\documentclass[11pt]{article} 

\usepackage[utf8]{inputenc} 

\usepackage{geometry} 
\usepackage{graphicx} 
\usepackage[parfill]{parskip} 
\usepackage{booktabs} 
\usepackage{array} 
\usepackage{paralist} 
\usepackage{verbatim} 
\usepackage{subfig} 
\usepackage{authblk}
\usepackage{amssymb,amsmath}
\usepackage{color}

\usepackage{fancyhdr} 
\usepackage{sectsty}
\allsectionsfont{\sffamily\mdseries\upshape} 

\def\ifb{fb$^{-1}$}
\def\iab{ab$^{-1}$}

\def\kv{\kappa_V}
\def\kl{\kappa_\ell}
\def\ku{\kappa_u}
\def\kd{\kappa_d}

\def\kg{\kappa_g}
\def\ka{\kappa_\gamma}

\def\sb{\sin\beta}
\def\cb{\cos\beta}
\def\sa{\sin\alpha}
\def\ca{\cos\alpha}

\def\nn{\nonumber}

\newcommand{\be}{\begin{equation}}
\newcommand{\ee}{\end{equation}}
\newcommand{\bi}{\begin{itemize}}
\newcommand{\ei}{\end{itemize}}
\newcommand{\bea}{\begin{eqnarray}}
\newcommand{\eea}{\end{eqnarray}}

\title{Scrutinizing $h(125)$ in Two Higgs Doublet Models at the LHC, ILC, and Muon Collider}
\author[1]{Vernon~Barger\thanks{barger@pheno.wisc.edu}}
\author[1]{Lisa~L.~Everett\thanks{leverett@wisc.edu}}
\author[2]{Heather~E.~Logan\thanks{logan@physics.carleton.ca}}
\author[1]{Gabe~Shaughnessy\thanks{gshau@hep.wisc.edu}}
\affil[1]{Department of Physics, University of Wisconsin, Madison, WI 53706, USA}
\affil[2]{Ottawa-Carleton Institute for Physics, Carleton University, Ottawa, ON K1S 5B6, Canada}

\date{} 

\begin{document}

\maketitle

\begin{abstract} 
The discovery at the LHC of a scalar particle with properties that are so far consistent with the SM Higgs boson is one of the most important advances in the history of particle physics.  The challenge of future collider experiments is to determine whether its couplings will show deviations from the SM Higgs, as this would indicate new physics at the TeV scale, and also to probe the flavor structure of the Yukawa couplings.    As a benchmark alternative to the SM Higgs, we consider a generic two Higgs doublet model (2HDM)  and analyze the precision to which the LHC14, an ILC250, 500, 1000 GeV and a 125 GeV Muon Collider (MC) can determine the gauge and Yukawa couplings.  We allow for correlations among the couplings.   We include the impact of a Higgs total width measurement, indirectly at the LHC and ILC and by a direct scan at the MC.   We also discuss pattern relations among the couplings that can test for singlet or doublet Higgs extensions of 2HDMs.

\end{abstract}

\newpage

\section{Introduction}
\label{sec:intro}

Particle physics is at a crossroads.  The long-sought Higgs boson has been found at the LHC, confirming the brilliant prediction that the electroweak gauge symmetry of the Standard Model (SM) gauge theory must be spontaneously broken.  This monumental advance is tempered by the absence of evidence at the LHC for new physics that could explain the large hierarchy between the electroweak and Planck scales, which is destabilized by quadratic radiative corrections to the Higgs mass and would require extreme fine-tuning in the absence of new physics. As such fine-tuning runs counter to the philosophy that theory should explain fundamental physics, the SM is presumed to be incomplete.  Supersymmetry (SUSY) provides a way out of this conundrum through cancellation of the quadratic SM radiative corrections by corresponding loops involving ``superpartner" particles. The absence of LHC evidence for the colored SUSY particles, the squarks and the gluino, has excluded the masses of these particles below the TeV scale and begins to strain naturalness criteria, wherein the necessary cancellations among SM and SUSY contributions to the light-Higgs mass and the $Z$-boson mass should not be fine-tuned (e.g. Ref.~\cite{Baer:2012cf} and references therein).  However, it is found to be possible to maintain electroweak naturalness with first and second generations quarks of $10-20$ TeV masses, which also ameliorates a SUSY problem with flavor physics~\cite{Gabbiani:1996hi,Arvanitaki:2012ps}.  

Within the minimal supersymmetric standard model (MSSM),  the Higgs sector consists of two doublets. In the absence of CP violation, the physical Higgs states are two CP even states, $h$ and $H$, a CP odd state, $A$, and charged Higgs states, $H^\pm$.  For an $H$ mass that is much larger than the $h$ mass, the light Higgs becomes SM-like, the so-called decoupling limit~\cite{Gunion:2002zf}.  The existence of $H$ causes modifications in the $h$ couplings, due to their mixing, albeit the corrections are small for heavy $H$.  It is these modifications to the $h$ couplings that would signal the existence of new physics, as discussed recently e.g.~in~\cite{Branco:2011iw,Gupta:2012mi,Killick:2013mya}.  

Supersymmetry is one of several theoretical models that can tame the quadratic divergences of the Higgs sector or push the UV problem to higher energies.   Others proposals include the Little Higgs model and its variants which introduce heavy top-like states and models with extra-dimensions.  SUSY is nominally deemed to be the most compelling model, because it explains the convergence of the strong, weak and electromagnetic couplings at the GUT scale as well as the quantum number assignments of the quarks and leptons in 16 dimensional representations of a $SO(10)$ GUT gauge group.  Additionally, SUSY provides a WIMP candidate for the cold dark matter (CDM) in the Universe, in addition to the SM CDM axion particle that can solve the strong CP problem. However, as the SUSY particles still await discovery, it is prudent to be open-minded about the eventual theory.  Thus, in exploring possible deviations in the Higgs couplings from those of the SM Higgs, we choose a generic 2HDM as a benchmark for comparison of how future colliders can do in detecting deviations from the SM couplings.

Since the light Higgs has been found, it makes good sense to study its properties to the fullest extent possible at future colliders.  It is still a possibility, though this may seem remote from the arguments above in favor of supersymmetry, that the particle could be the SM Higgs, a thought that is being entertained.  If that is the case, then the Universe seemingly resides in an unstable minimum but with a lifetime that exceeds the present age of the Universe~\cite{Buttazzo:2013uya}.  

It has been conjectured that the SM Higgs mass may be predicted by the vanishing of the Higgs self-coupling and its beta function in the vicinity of the Planck scale, but a solid theoretical justification of these conditions is missing.  Also, renormalization group evolution of the self-coupling with present data seems to show that it changes sign at an intermediate scale, ${\cal O}(10^{10})$ GeV\cite{Buttazzo:2013uya}.  The SM Higgs scenario also mandates precision studies of the light Higgs couplings~\cite{Cheung:2013kla,Gainer:2013rxa}. 

New physics may also be probed at the energy frontier, through the direct production of any new physics particles.  For example, the discovery of a heavier neutral Higgs, $H$ or $A$, or a charged Higgs, would immediately show the necessity of an extended Higgs sector.  The discovery of squarks and gluinos at LHC14 would likewise point to the existence of additional Higgs states~\cite{Baer:2012vr}.  Low energy measurements, such as rare $B$-decays, that could be found to deviate from SM predictions, are another probe (c.f. Ref.~\cite{Buras:2013ooa}).  These are complementary to a precision study of the light Higgs that we pursue here.

As mentioned above, it is useful to introduce a benchmark new physics model  for comparisons with future data and we select a generic 2HDM for this purpose~\cite{Pich:2009sp,Cree:2011uy,Altmannshofer:2012ar,Bai:2012ex} that does not conflict with other associated new physics that may or may not exist.  The model is chosen such that the couplings of the quarks to the Higgs doublets are aligned to eliminate dangerous flavor-changing neutral currents at tree-level.  The scenario encompasses the traditional types of 2HDMs, namely Type I, Type II, lepton-specific, and flipped models; the SUSY model is Type II.  The model provides an encompassing framework to analyze forthcoming data.  Our study is independent of new physics that may be uncovered in other ways.  Previous studies on various 2HDMs in light of the LHC Higgs discovery may be found in Refs.~\cite{Akeroyd:1996he,Ferreira:2011aa,Blum:2012kn,Cheon:2012rh,Carmi:2012in,Ferreira:2012nv,Chang:2012zf,Chen:2013kt,Celis:2013rcs,Giardino:2013bma,Grinstein:2013npa,Shu:2013uua,Barroso:2013zxa,Coleppa:2013dya,Eberhardt:2013uba,Craig:2013hca}.  The model does not in itself solve the quadratic divergence problem of the SM, for which additional new physics is also needed.

There are two ways that the 2HDM can be used in analyses of future data on the light Higgs with machine-specific anticipated measurement uncertainties as inputs. One is to exclude regions of modified couplings, which is the approach of our current study.   The other is to assume a particular measured value and determine how well it can be distinguished from the SM value.  Although the latter gives a positive slant in the discovery of new physics effects, it requires advance knowledge of the couplings for which the new physics may be manifest.  In our study we make no assumptions about the origin of loop corrections that lead to the observed light Higgs boson mass.  We treat all couplings at tree-level, which is reasonable for the light Higgs since its tree-level couplings are not small.  We do not consider triplet or larger Higgs representations in our comparative study~\cite{Triplets}, nor do we explicitly consider the constraints of the heavy Higgs scalars,  for which a treatment has been recently given~\cite{Chen:2013rba}.  Our analysis does take into account the multi-dimensional parameter space of couplings and their simulated experimental correlations.

In Section~\ref{sect:fit}, we present the fit technique we adopt.  In Section~\ref{sect:genfit}, we present the results of fit to the LHC and Tevatron data assuming a general model with freely varying Higgs couplings, while in Section~\ref{sect:2hdms}, we show the fit for the various flavors of 2HDMs.  In Section~\ref{sect:width}, we discuss the added benifit of measuring the Higgs width, while in Section~\ref{sect:pattern}, we show how one may distinguish the mixing of additional $SU(2)$ doublets and singlets.  Finally, in Section~\ref{sect:concl}, we conclude and summarize our results. 

\section{Simulation and Fit Technique}
\label{sect:fit}

To arrive at the model predictions, we utilize a Markov Chain Monte Carlo (MCMC) approach that is efficient for scanning parameters with high dimensionality.  While the MCMC approach is Bayesian in nature, we extract the 1$\sigma$ and 95\% C.L. regions consistent with the Frequentist approach.  In this regard we are simply using the MCMC as a tool to optimally search the parameter space for points lying near the bottom of the $\chi^2$ profile.  We determine $\chi^2_{\rm min}$, the minimum $\chi^2$ for each scenario, then define the 1$\sigma$ and 95\% C.L. regions with a $\Delta \chi^2 = \chi^2 - \chi^2_{\rm min} = 2.3$ and 6.0, respectively, for a two parameter fit.

\begin{figure}[htbp]
\begin{center}
     \includegraphics[angle=0,width=0.5\textwidth]{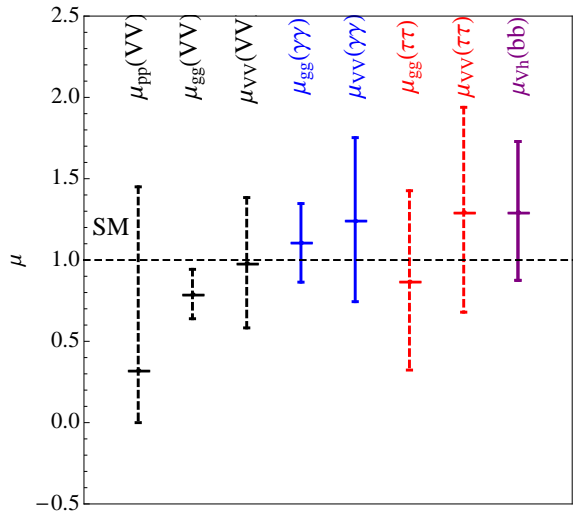}
\caption{Distilled measurements of the Higgs production cross sections with respect to the SM from the Tevatron and LHC.  The horizontal line at $\mu=1$ denotes the SM expectation. Color coding is grouped by decay mode.}
\label{fig:pullplot}
\end{center}
\end{figure}

We fit the LHC data as of August 2013 following the methodology of Ref.~\cite{Low:2012rj}, and have verified it provides good agreement with the fits from the ATLAS and CMS collaborations~\cite{ATLAS:2012wma,CMS:yva}.  Other global Higgs fits have been made in the literature, but with different methodology and data inputs~\cite{Belanger:2013xza}.  Our method involves separately varying $\kv,\ku,\kd$ and $\kl$, the parameters that modify the SM Vector Boson and Yukawa couplings, respectively.  From these coupling scalings, the values of $\ka$ and $\kg$ can be calculated within the 2HDM.  We assume a sufficiently heavy charged Higgs such that the $\gamma\gamma$ rate is not significantly affected by it.  In Appendix~\ref{apx:CurrentHdata}, we show the values included in our fit from Tevatron~\cite{Tevatron:2012cn}, CMS~\cite{CMS:yva,CMS:ril,CMS:13015,CMS:bxa,CMS:eya,CMS:zwa,CMS:xwa,CMS:utj,CMS:ckv,CMS:13011,CMS:13012} and ATLAS~\cite{ATLAS:2013oma,ATLAS:2013wla,ATLAS:2013nma,ATLAS:2012dsy,ATLAS:2012wma} up to August-2013.  These data can be distilled into the following measurements on $\mu_{\rm prod}(X\bar X)$, the production cross section of the Higgs boson decaying to $X\bar X$ scaled with respect to the SM:
\bea
\mu_{p\bar p}(\gamma\gamma)&=& 3.62^{+2.96}_{-2.54} \,, \quad
\mu_{gg}(\gamma\gamma)= 1.1^{+0.2}_{-0.2} \,, \nn\\
\mu_{VV}(\gamma\gamma)&=& 1.2^{+0.5}_{-0.5}   \,, \quad 
\mu_{gg}(VV)= 0.79^{+0.15}_{-0.15} \,, \\
\mu_{VV}(VV)&=& 1.0^{+0.4}_{-0.4} \,,  \quad
\mu_{gg}(\tau\tau)= 0.9^{+0.6}_{-0.5}\,, \nn\\
\mu_{VV}(\tau\tau)&=& 1.3^{+0.6}_{-0.6} \,, \quad
\mu_{Vh}(b\bar b) = 1.3^{+0.4}_{-0.4} \,,\nn
\eea
where the low precision Tevatron measurement of $\mu_{pp}(\gamma\gamma)$ is from for the inclusive $h\to \gamma\gamma$ rate. These  measurements are compared with their SM expectations in Fig.~\ref{fig:pullplot}.  

We include projected LHC sensitivities at $\sqrt s=14$ TeV for 300
fb$^{-1}$ and 3 ab$^{-1}$, denoted as LHC300 and LHC3000,
respectively, according to the uncertainties on signal strengths given in
Ref.~\cite{ATLAS:2013hta,CMS:2013xfa}.  For the ATLAS projections, we
include the theory systematics~\cite{ATLAS:2013hta}.  For the
CMS projections, we use ``Scenario 2'', which extrapolates the
analyses of 7 and 8 TeV data to 14 TeV  assuming the theory
uncertainties will be reduced by a factor of 2 while other uncertainties
are reduced by a factor of $1/\sqrt{\cal L}$~\cite{CMS:2013xfa}.

Since the CMS signal-strength projections are not separated by production mode, we assume inclusive production for all decay modes except $b \bar b$, which we assume proceeds via $W/Z$-Higgsstrahlung.  This neglects the preferential weighting of clean but rare signal event categories in the multivariate signal strength analyses when high statistics are available. It also means that we lose the information on the relative strengths of different Higgs production modes that is available in a full analysis but is not provided by CMS~\cite{CMS:2013xfa}. We show in Table~\ref{tab:lhcunc} the LHC uncertainties that we adopt for the two luminosity benchmarks.

The CMS collaboration also provides projections for the uncertainties on seven individual Higgs couplings $\kappa_W$, $\kappa_Z$, $\kappa_u$, $\kappa_d$, $\kappa_{\ell}$, $\kappa_{\gamma}$, and $\kappa_g$.  However, implementing these directly into our fit neglects the important correlations among the couplings.  Ideally, the experiments would provide the full covariance matrix for this coupling extraction.  Since this information is not public, we choose to proceed using the signal strengths as described above.  We remark that our fit yields uncertainties on the Higgs couplings $\kappa_V$, $\kappa_u$, $\kappa_d$, and $\kappa_{\ell}$ (see Table~\ref{tab:fit}) that are slightly larger than those quoted by CMS for their seven-coupling extraction~\cite{CMS:2013xfa}.

\begin{table}[htdp]
\caption{Projected $1\sigma$ sensitivities by channel for the LHC operating at $\sqrt s=14$ TeV.  The 300 fb$^{-1}$  and 3 ab$^{-1}$ uncertainties are taken from Refs.~\cite{CMS:2013xfa,ATLAS:2013hta}.  See text and Ref.~\cite{CMS:2013xfa,ATLAS:2013hta} for further details.}
\begin{center}
\begin{tabular}{|cc|cc|}\hline
Experiment & Channel & 300 fb$^{-1}$ & 3 ab$^{-1}$\\
\hline
ATLAS & $\mu_{gg}(\gamma\gamma)$  & 15\% & 13\%\\
ATLAS &  $\mu_{gg}(\gamma\gamma+j)$  & 16\% & 12\%\\
ATLAS &  $\mu_{VV}(\gamma\gamma)$  &  34\%& 16\%\\
ATLAS &  $\mu_{VH}(\gamma\gamma)$  &  77\%& 25\%\\
ATLAS &  $\mu_{t\bar t}(\gamma\gamma)$  &  55\%& 21\%\\
ATLAS  & $\mu_{gg}(ZZ)$  & 16\% & 13\%\\
ATLAS  & $\mu_{gg}(WW)$  &  29\% & 29\%\\
ATLAS &  $\mu_{VV}(WW)$  &67\%  & 58\%\\
ATLAS &  $\mu_{gg}(\mu^+\mu^-)$  & 53\% & 21\%\\
ATLAS &  $\mu_{t\bar t}(\mu^+\mu^-)$  & 73\% & 26\%\\
CMS &  $\mu_{pp}(\gamma\gamma)$  & 6\% & 4\%\\
CMS &  $\mu_{pp}(WW)$  & 6\% & 4\%\\
CMS &  $\mu_{pp}(ZZ)$  & 7\% & 4\%\\
CMS &  $\mu_{VH}(b\bar b)$  & 11\% & 5\%\\
CMS &  $\mu_{pp}(\tau^+\tau^-)$  & 8\% & 5\%\\
CMS &  $\mu_{pp}(Z\gamma)$  & 62\% & 20\%\\
\hline
\end{tabular}
\end{center}
\label{tab:lhcunc}
\end{table}%

The coupling uncertainties at an ILC operating at 250, 500 and 1000 GeV with integrated luminosities of 250, 500 and 1000 fb$^{-1}$, respectively, are obtained by each experiment from the cross section uncertainties for the various final states in~\cite{Baer:2013cma}.  The beam polarizations can be tuned to emphasize  the physics process in question.  For Higgs measurements, at 250 and 500 GeV, the beam polarizations are assumed to be $(e^-,e^+)=(-0.8,+0.3)$, while at 1 TeV, the polarization is  $(e^-,e^+)=(-0.8,+0.2)$.  At 250 GeV, the ILC will be well positioned to produce Higgs bosons copiously and therefore will obtain a direct measurement of the total Higgs production rate via $Z$ Higgsstrahlung to an accuracy of $\approx 2.5\%$~\cite{Barger:1992ei,Baer:2013cma}.  With higher energies, the Vector Boson fusion cross section grows, and becomes the dominant Higgs production channel above $\sqrt s \gtrsim 400$ GeV~\cite{Baer:2013cma}.  Therefore, at $\sqrt s = $ 500 GeV and 1 TeV, an independent accurate measurement of the Vector Boson coupling strength can be made.  We show these uncertainties in Table~\ref{tab:ilcunc}.
\begin{table}[htdp] 
\caption{Projected sensitivities by channel for the ILC operating at $\sqrt s=250$, 500 and 1000 GeV with a corresponding integrated luminosity of ${\cal L}~dt=250$, 500 and 1000 fb$^{-1}$, respectively.  From Ref.~\cite{Peskin:2012we,Baer:2013cma}.}
\begin{center}
\begin{tabular}{|c|ccc|}\hline
Channel & 250 GeV & 500 GeV &  1 TeV\\
\hline
$\mu_{Zh}$  & 2.5\% & -- & --\\
$\mu_{Zh}(b\bar b)$  &  1.1\% & 1.8\% & --\\
$\mu_{Zh}(c\bar c)$  & 7.4\% & 12\% & --\\
$\mu_{Zh}(gg)$ &  9.1\%  & 14\% & --\\
$\mu_{Zh}(WW)$ &6.4\%   & 9.2\% & --\\
$\mu_{Zh}(ZZ)$  & 19\% & 25\% & --\\
$\mu_{Zh}(\tau\tau)$  & 4.2\%  & 5.4\% & --\\
$\mu_{Zh}(\gamma\gamma)$  & 38\%  & 38\% & --\\
$\mu_{WW}(b\bar b)$  &  11\% & 0.66\% & 0.47\%\\
$\mu_{WW}(c\bar c)$  &  -- & 6.2\% & 7.6\%\\
$\mu_{WW}(gg)$  &  -- & 4.1\% & 3.1\%\\
$\mu_{WW}(WW)$  &  -- & 2.6\% & 3.3\%\\
$\mu_{WW}(ZZ)$  &  -- & 8.2\% & 4.4\%\\
$\mu_{WW}(\tau\tau)$  &  -- & 14\% & 3.5\%\\
$\mu_{WW}(\gamma\gamma)$  &  -- & 26\% & 10\%\\
$\mu_{WW}(\mu\mu)$  &  -- & -- & 32\%\\
$\mu_{t\bar t}(b\bar b)$  &  -- & 25\% & 8.7\%\\
\hline
\end{tabular}
\end{center}
\label{tab:ilcunc}
\end{table}%

The Higgs width may be measurable at a $\gamma\gamma$ collider, with  uncertainty of 8\%~\cite{Ohgaki:1997jp}.  In a $\mu\mu$ collider, the accuracy on the total width can be improved to 3.6\% under default design assumptions~\cite{Han:2012rb}.  We discuss the impact of  the total Higgs width measurement on the determination of the 2HDM in section~\ref{sect:width}.

\section{General Fit}
\label{sect:genfit}
Before discussing the specific 2HDM that we will adopt, we present the general fit to simulated data.  We assume the  Higgs-like boson $h$ observed at the LHC is a linear combination of the neutral CP-even components of the two $SU(2)$ doublets $\Phi_{1,2}$, in which $\Phi_1$ has $Y=-1$ and $\Phi_2$ has $Y=1$.  We assume that the Vector Bosons couple to the Higgs boson in a way that is consistent with custodial symmetry:
\be
g_W = {g^2 v\over 2} \kv,\quad g_Z={\left(g^2+{g^\prime}^2\right)v \over 2} \kv,
\ee
To unitarize longitudinal Vector Boson scattering in the case of one Higgs doublet , $\kv=1$. Generally, with additional doublets or singlets, 
$\kv \le 1$.  However, larger representations may allow $\kv > 1$.  Unitarization in either case is achieved through the Higgs states.  Since we have no strong constraints on the heavy Higgs bosons that would unitarize scattering, we allow the value of $\kv$ to float in our fits.

Likewise, we assume each fermion has a coupling that is independent, but maintain generation universality, such that the neutral CP-even portion of the Yukawa Lagrangian is
\be
-{\cal L}_{\rm Yuk} = \ku y_u \bar u_R {h\over \sqrt 2} u_L+\kd  y_d \bar d_R  {h\over \sqrt 2} d_L+\kl  y_\ell  \bar\ell_R  {h\over \sqrt 2} \ell_L + {\rm h.c.},
\ee
where the SM Yukawa coupling is $y_f = \sqrt 2 m_f/ v$, where $v=246$ GeV.

The loop-induced gluon and photon couplings are given by the known functions $F_j$~\cite{Gunion:1989we}.  At leading order, the CP-even $h \gamma\gamma$ effective couplings are given by
\be
{\cal A} = {\alpha_{\rm em}\,M_{h}^2\over 4 \pi v_{\rm EW}} \sum_{j=q,\ell,W^\pm} N_{c j} Q_j^2 \kappa_{j} F_{j}(\tau_{j}). 
\ee
where for each loop particle $j, N_{cj}$ is the color factor, $Q_j$ is the charge, and $\tau_{ij}={4 m_j^2/M_{h}^2}$.  The gluon amplitude is similar, but it contains only the quark loop~\cite{Gunion:1989we}.  For the general model, we assume no further contributions to these loop induced couplings.\footnote{Exceptions to this include fourth generation chiral fermions that substantially alter the loop induced $hgg$ and $h\gamma\gamma$ couplings~\cite{Kribs:2007nz}.} 

We fit a cross section extracted from measurement with respect to the SM expectation by assuming
\be
\mu_{\rm prod}(X\bar X) = {\Gamma_{\rm prod}\over\Gamma^{\rm SM}_{\rm prod}}  {\Gamma_{X\bar X} \over \Gamma_{\rm total}} {\Gamma^{\rm SM}_{\rm total} \over \Gamma^{\rm SM}_{X\bar X}},
\ee
where the first factor accounts for the cross section scaling and other factors account for the modifications in the branching fraction of $h\to X\bar X$.  At the LHC, for the inclusive production modes, we sum over the gluon fusion and Vector Boson fusion contributions:
\be
\sigma_{pp\to h} =  \kg^2 \sigma_{gg\to h} + \kv^2 \sigma_{VV\to h}, \quad\quad {\Gamma_{pp}\over \Gamma^{\rm SM}_{pp}} = \kg^2 \epsilon_{gg} + \kv^2 (1-\epsilon_{gg}),
\ee
where $\epsilon_{gg}$ is the gluon fusion fraction of the inclusive rate.  We adopt the value for  $M_h=125$ GeV of $\epsilon_{gg}=0.92$ 
for both 7 TeV and 14 TeV~\cite{Dittmaier:2011ti}.  We neglect the $W/Z$-Higgsstrahlung process as it is quite small.  The scaling factors $\kg$ and $\kv$ account for the $gg$ and $WW$ or $ZZ$ couplings, respectively.

For Vector Boson fusion and $W/Z$-Higgsstrahlung initiated production,
\be
{\Gamma_{VV}\over \Gamma^{\rm SM}_{VV}} = {\Gamma_{Vh}\over \Gamma^{\rm SM}_{Vh}} = \kv^2.
\ee

\begin{figure}[htdp]
\begin{center}
     \includegraphics[angle=0,width=0.47\textwidth]{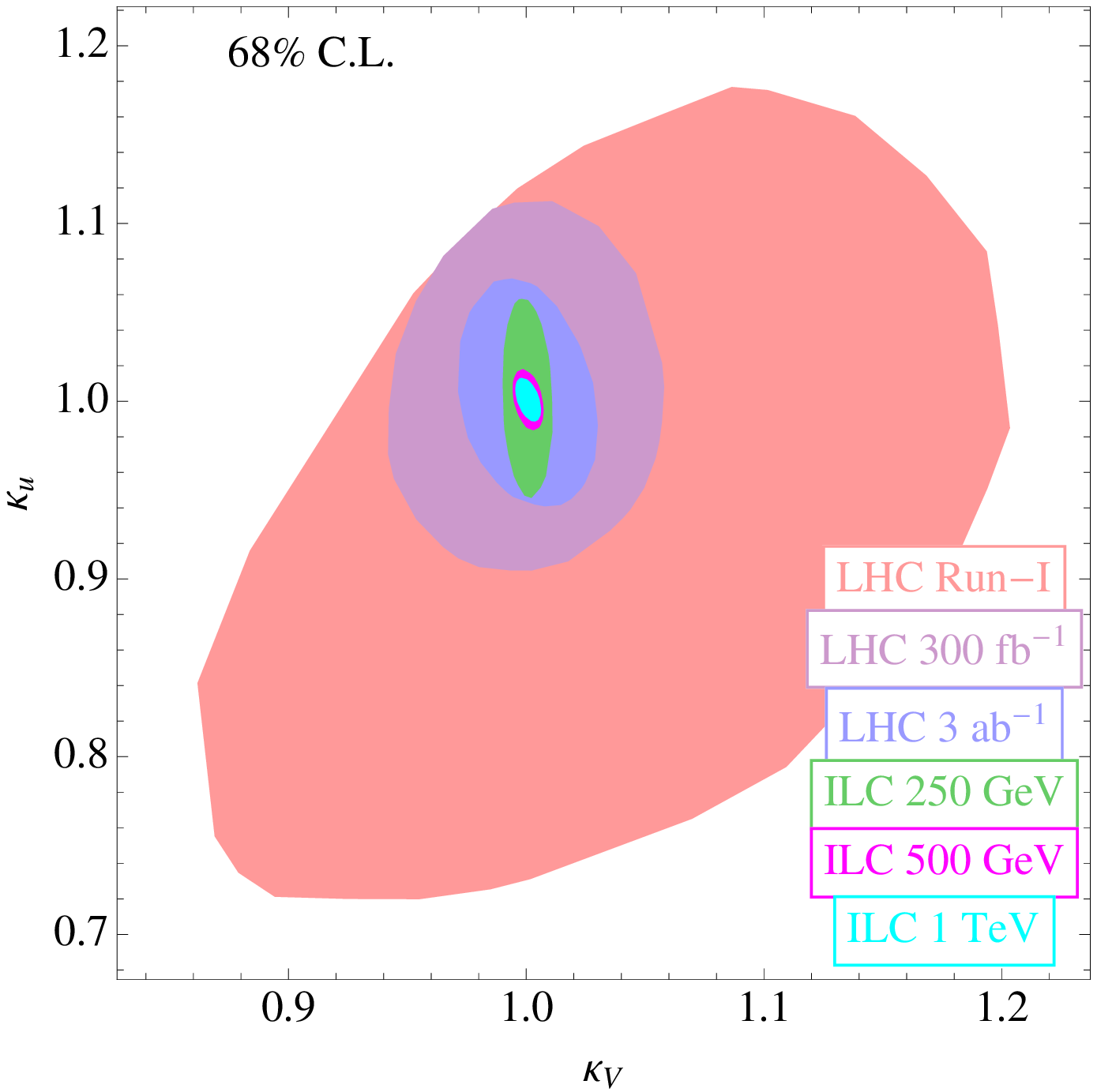}
     \includegraphics[angle=0,width=0.47\textwidth]{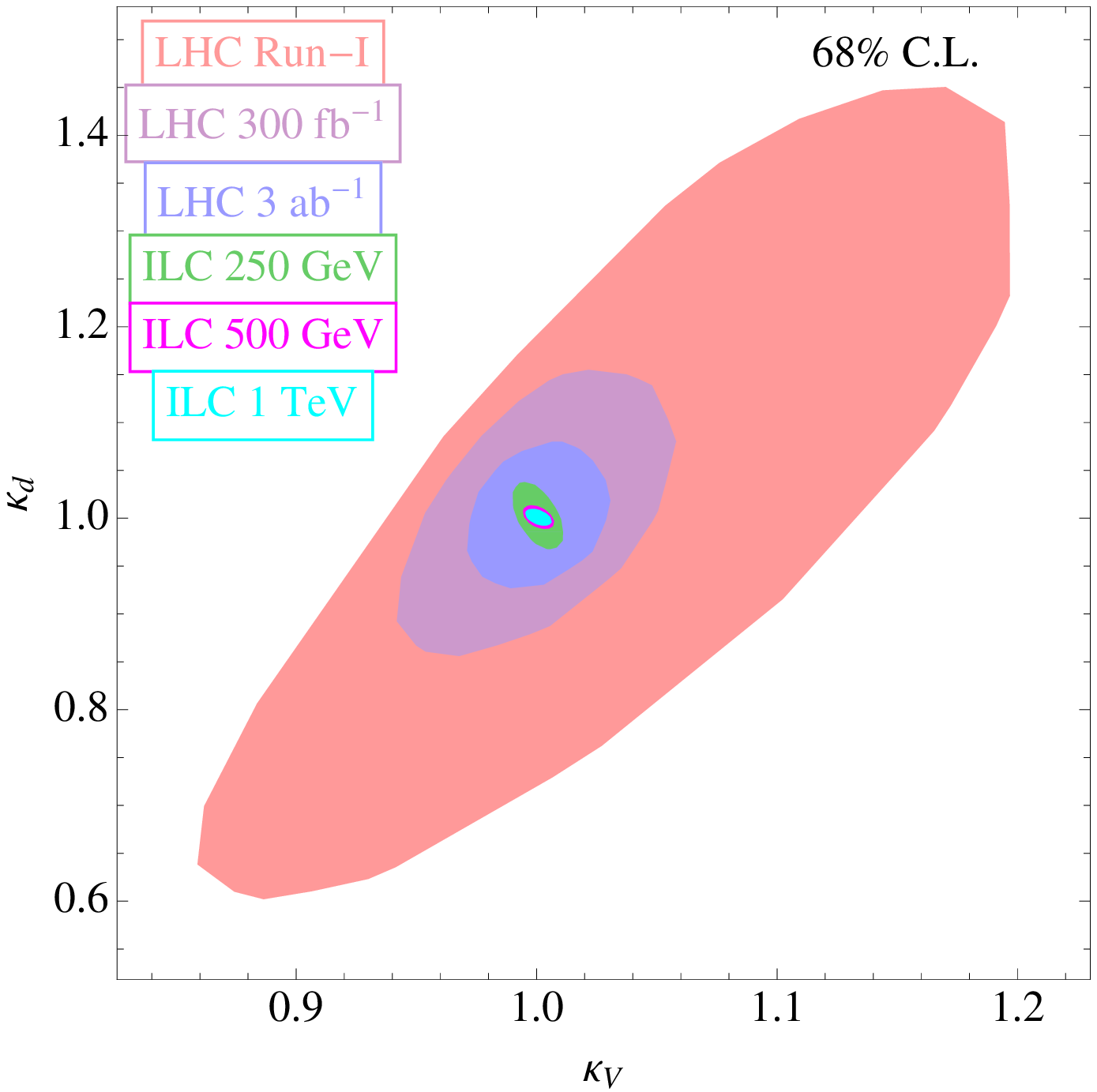}\\
     \includegraphics[angle=0,width=0.47\textwidth]{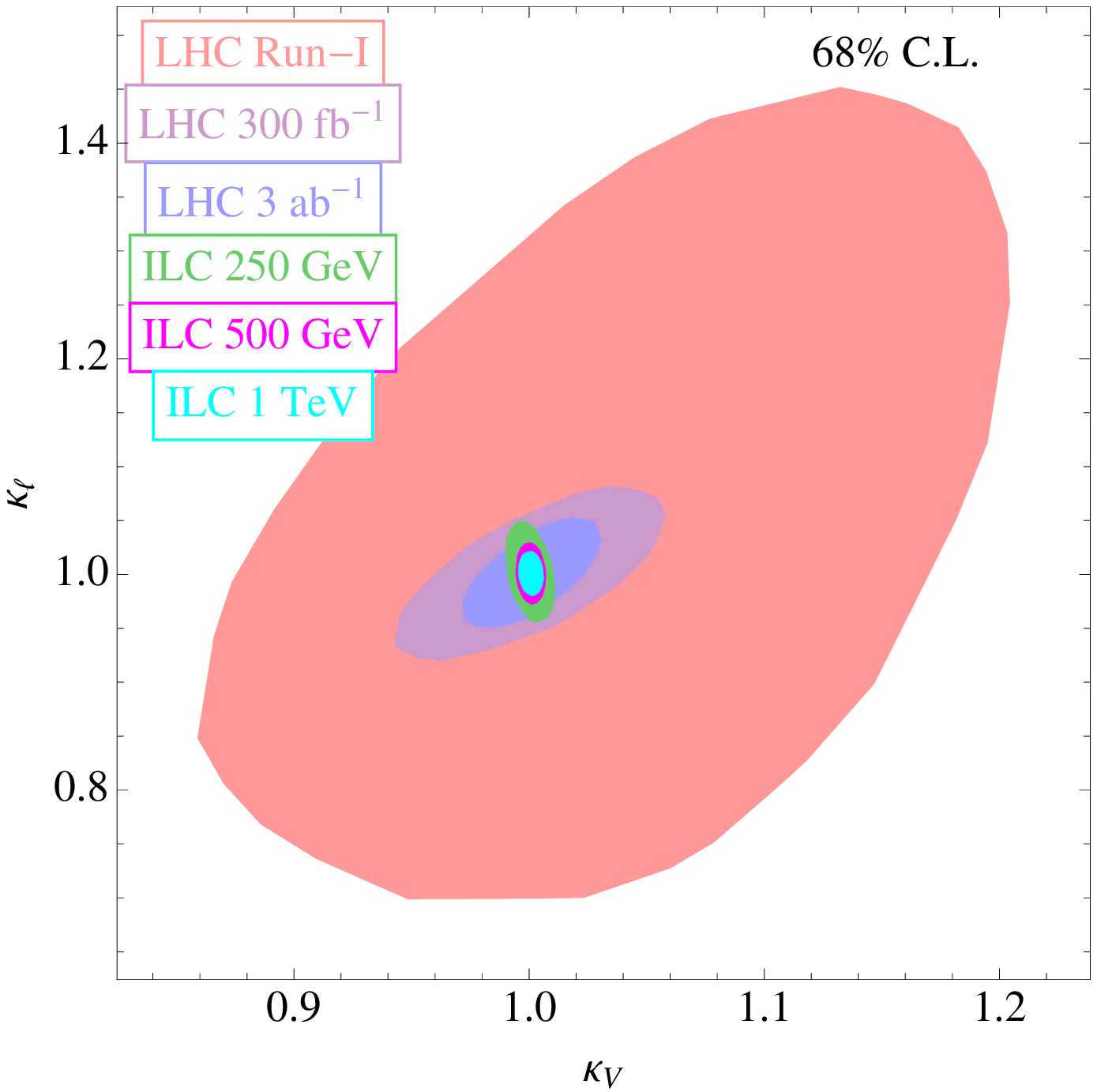}
     \includegraphics[angle=0,width=0.47\textwidth]{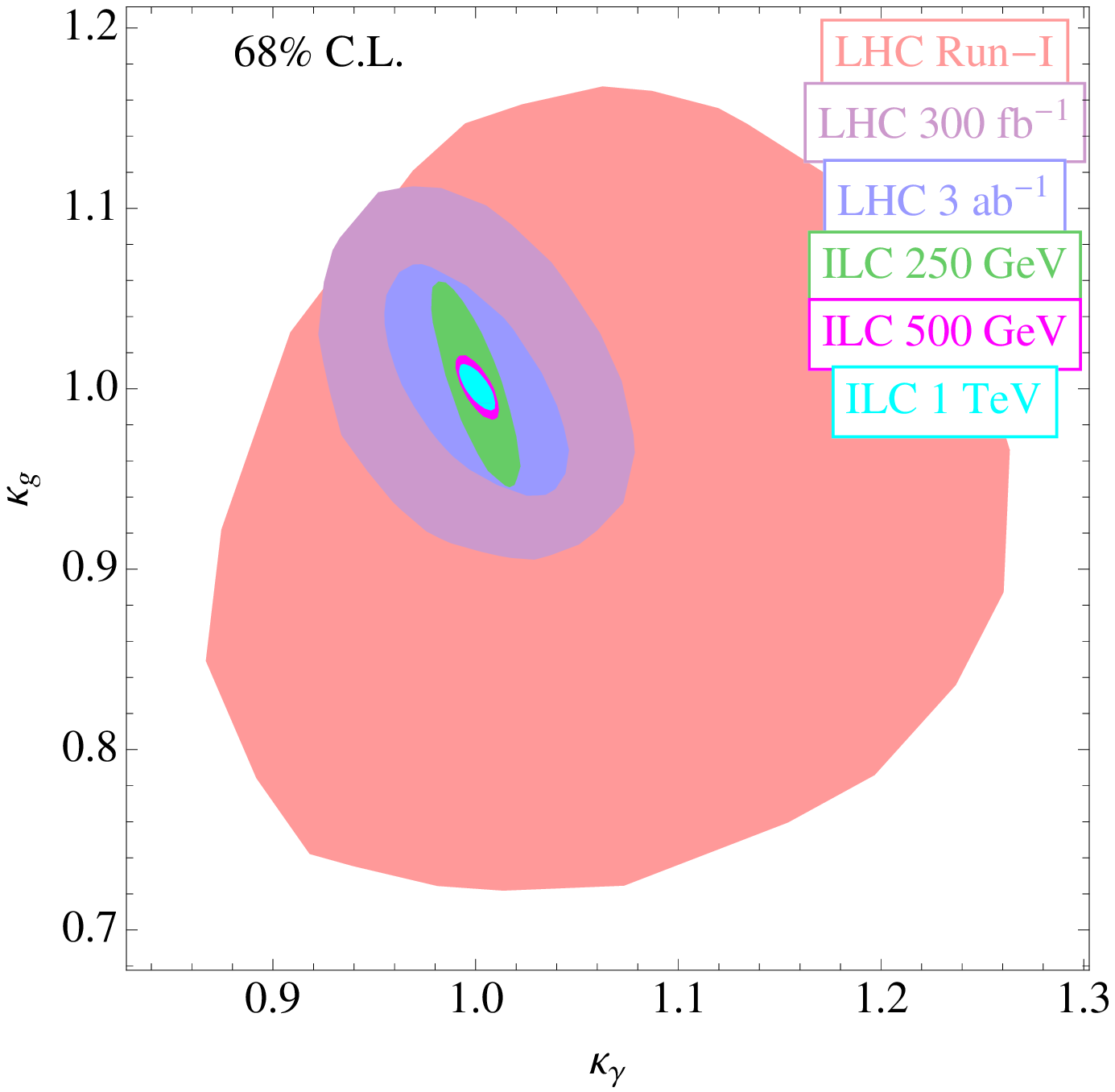}
\caption{General fit to current data (as of August 2013) and future LHC and ILC data at the 1$\sigma$ level.  The ILC operating at 250, 500 and 1000 GeV is assumed to collect to 250, 500 and 1000 fb$^{-1}$ of integrated luminosity, respectively }
\label{fig:genfit}
\end{center}
\end{figure}

\begin{table}[htdp]
\caption{Coupling sensitivities of current LHC and future LHC \& ILC data.  Uncertainties are obtained via a 1d fit and are given at the $1\sigma$ level; those quoted as a \% are symmetric.  The LHC-I (Unc.) column gives the uncertanties as a \% of the central value in the LHC-I column.}
\begin{center}
\begin{tabular}{|c|cc|cc|ccc|}
\hline
 	& LHC-I  & LHC-I  (Unc.) & LHC300  & LHC3000 & ILC250 & ILC500 & ILC1000\\
\hline
$\kv$ 	& $1.03\pm0.12$ &12\% &  4\%&  2\%&  0.7\%&  0.4\%&  0.4\%\\
$\ku$	&  $0.91^{+0.17}_{-0.15}$ & $^{+0.19}_{-0.16}$ &  8\%&  5\%&  4\%&  1.1\% &  0.8\%\\
$\kd$ 	&  $0.99^{+0.30}_{-0.29}$&$^{+0.30}_{-0.29}$&  11\%&  5\%&  2\%&  0.8\%&  0.6\%\\
$\kl$ 	&  $1.06^{+0.26}_{-0.26}$& $^{+0.25}_{-0.25}$&  6\% &  4\%  &  3\%&  2\%&  1\% \\
\hline
$\kd/\ku$	&  $1.09^{+0.19}_{-0.22}$&$^{+0.17}_{-0.20}$&  6\% &  3\%  &  3\%  &  1.3\%&  1.0\%\\ 
$\kl/\ku$	&  $1.15^{+0.31}_{-0.27}$&  $^{+0.27}_{-0.23}$&  9\% &  7\%&  4\% &  2\%&  1.5\%\\
\hline
\end{tabular}
\end{center}
\label{tab:fit}
\end{table}%

In Fig.~\ref{fig:genfit}, we show the 1$\sigma$ allowed regions in the plane of selected couplings, for the various collider assumptions as given above.   In addition to the fit of the current LHC results, we perform a fit to simulated future data under the assumption that no deviation from the SM expectation will be found.  The coupling uncertainties are generally ${\cal O}(10\%)$ or smaller as seen in Table~\ref{tab:fit}.  We consider points where the underlying model lies outside these regions to be more than $1\sigma$ from the SM result.   In these scenarios, we consider only the region in which $\ku,\kd,$ and $\kl \ge 0$.  While Yukawa couplings of opposite sign are possible, the top quark Yukawa is typically taken to be positive. In this case, any destructive interference occurs from $\kd$ and $\kl$, which imparts a shift of only ${\cal O}(0.1\%)$ to the coupling.

We note that the LHC Run-1 data prefer the $\ka>1$ region due to the higher $\gamma\gamma$ rate. This feeds into the estimation of $\ku$, which is preferred to be $\ku < 1$ to suppress of the destructive interference in the $t$ and $W$ mediated $h\to \gamma\gamma$ loops.  This is also somewhat afforded by the slightly lower $\kg$ value via the $gg\to h\to VV$ rate.

\begin{figure}[!t]
\begin{center}
     \includegraphics[angle=0,width=0.47\textwidth]{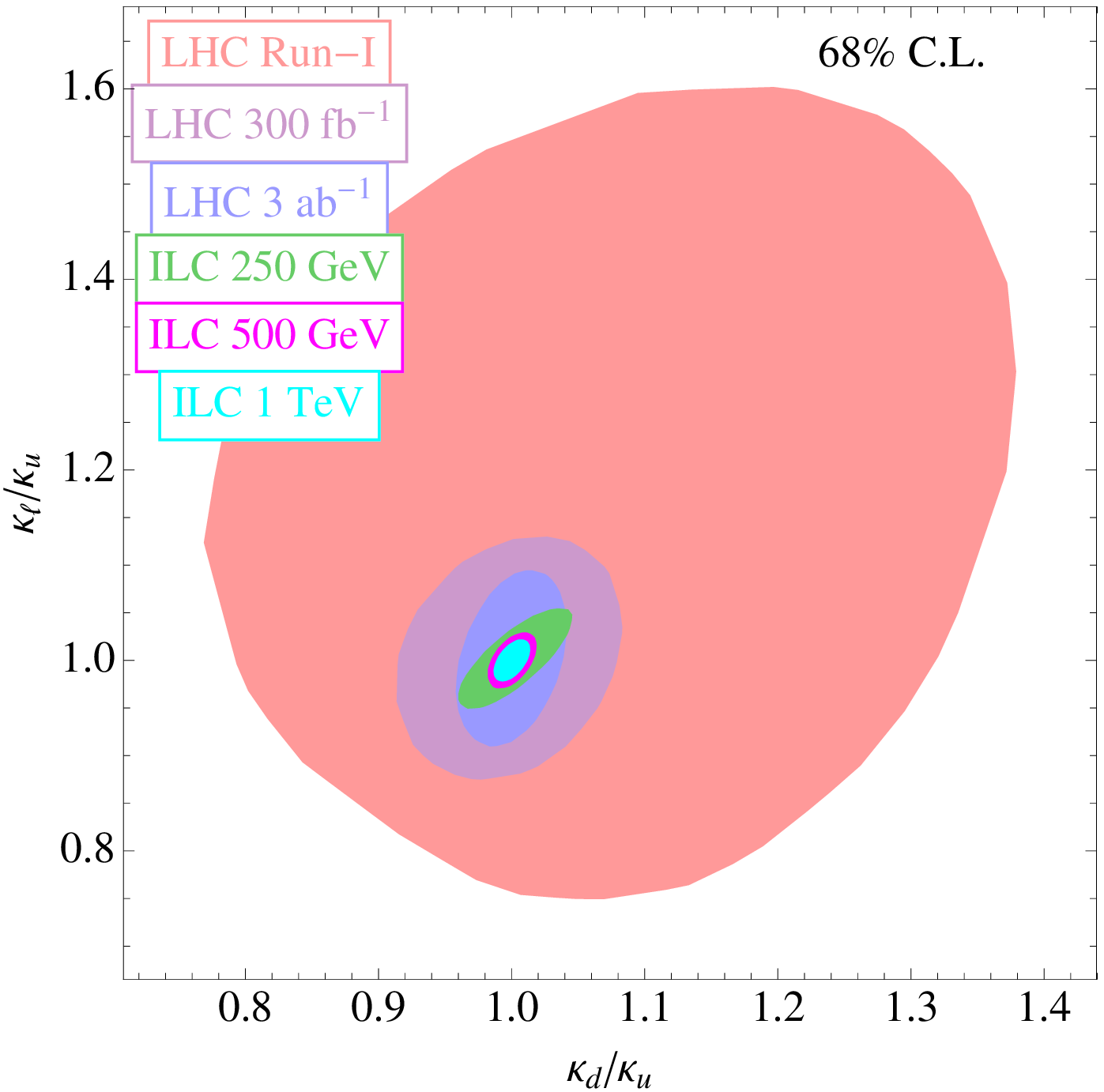}
     \includegraphics[angle=0,width=0.47\textwidth]{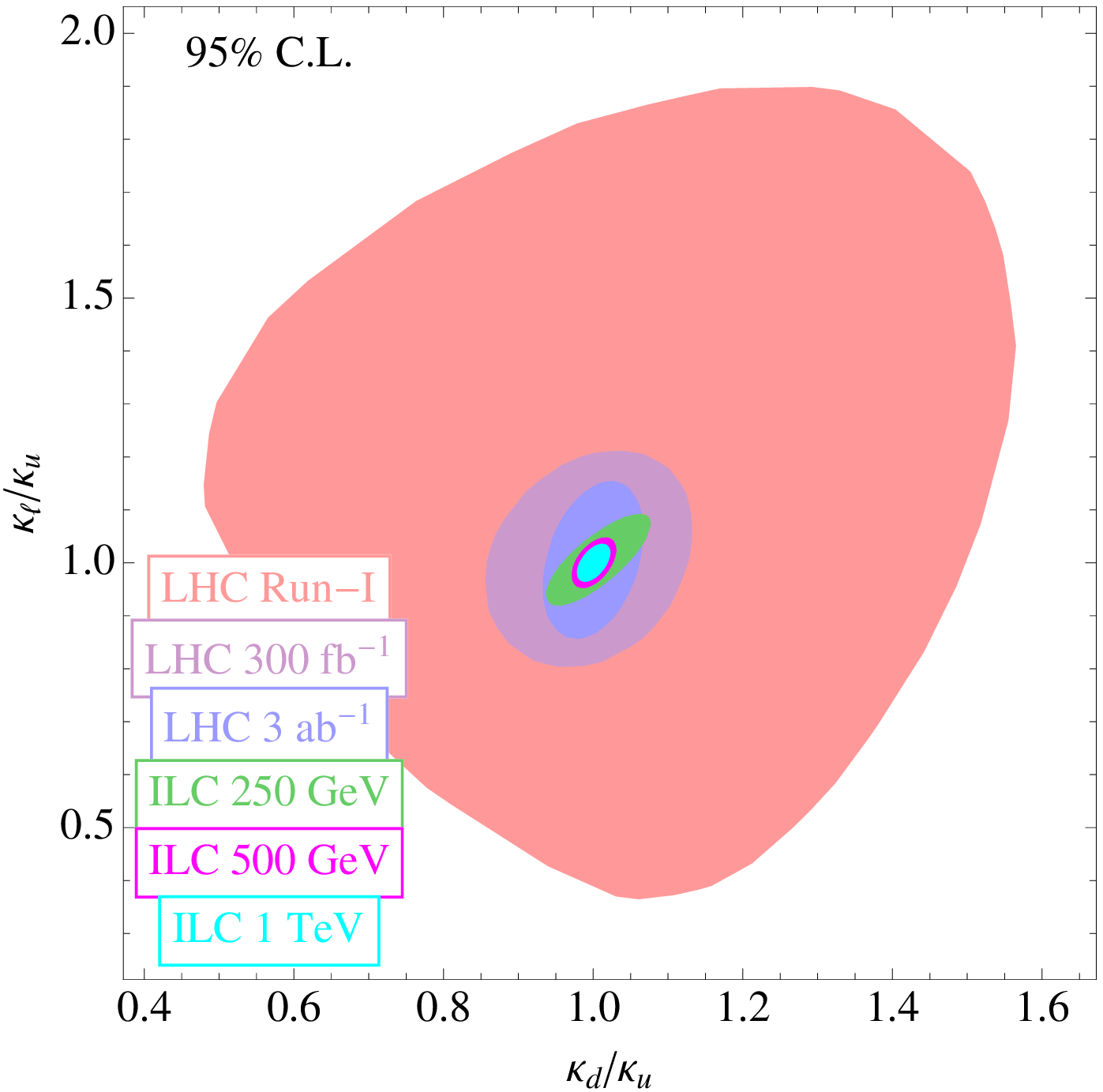}
\caption{Yukawa coupling ratios at the 1$\sigma$ (left panel) and 95\% C.L. (right panel) from a fit to the current data and simulated future LHC14 and ILC data.  The ILC operating at 250, 500 and 1000 GeV corresponds to 250, 500 and 1000 fb$^{-1}$ of integrated luminosity, respectively.  While $\kg$ and $\ka$ are derived parameters, their projected regions still are of interest.}
\label{fig:genfitdiff}
\end{center}
\end{figure}
In Fig.~\ref{fig:genfitdiff}, we show the Yukawa coupling ratios, $\kd/\ku$ and $\kl/\ku$, associated with each collider at 1$\sigma$ and 95\% C.L.  Deviations from the SM $(1,1)$ indicate a model that has non-trivial flavor structure.  We will discuss a class of 2HDMs that have these features.

\section{Two Higgs Doublet Models}
\label{sect:2hdms}
We work within a 2HDM framework that allows both doublets to couple to down quarks and charged leptons with aligned couplings~\cite{Pich:2009sp,Cree:2011uy,Altmannshofer:2012ar,Bai:2012ex}. As previously mentioned, the doublets are denoted by $\Phi_1$, which has $Y=-1$, and $\Phi_2$, which has $Y=1$.  With this notation in hand, we adopt the following parameterization of the Yukawa Lagrangian:
\bea
- {\cal L} &=& y_u\,\overline{u}_R \, {\bf \Phi}_2 \,Q_L    +\,y_d\, \overline{d}_R\,(\cos{\gamma_d}\,{\bf \Phi}_1 \,+\, \sin{\gamma_d}\,\tilde{{\bf \Phi}}_2) \, Q_L   \nonumber \\
&&\hspace{-3mm}+ \, y_\ell\,\overline{e}_R \, (\cos{\gamma_\ell}\,{\bf \Phi}_1 \,+\, \sin{\gamma_\ell}\,\tilde{{\bf \Phi}}_2) \,L_L     \,+\, \mbox{h.c.}\,,
\eea
in which  ${\bf \Phi}_{1,2}\equiv - {\bf \Phi}_{1,2}^* i \sigma_2$, and $\gamma_{\ell}$ and $\gamma_d$ parameterize the two Higgs doublet couplings to charged leptons and down quarks, respectively.  This general model can be mapped to the traditional 2HDMs via the angles specified in Table~\ref{tab:match}.  
\begin{table}[h!]
\renewcommand{\arraystretch}{1.5}
\centerline{
\begin{tabular}{c|cccc}
\hline \hline
Type     \hspace{2mm}         &     \hspace{1mm}   I          \hspace{2mm}      &  II    \hspace{2mm}  & L     \hspace{2mm}  &  Flipped   \hspace{2mm}  \\ \hline
$\gamma_\ell$  \hspace{2mm}  &  \hspace{1mm}  $\frac{\pi}{2}$  \hspace{2mm}  & 0  \hspace{2mm}   & 0  \hspace{2mm}   &   $\frac{\pi}{2}$  \hspace{2mm}  \\ \hline
$\gamma_d$  \hspace{2mm} & \hspace{1mm}      $\frac{\pi}{2}$          \hspace{2mm}  &  0 \hspace{2mm}  & $\frac{\pi}{2}$  \hspace{2mm}   & 0  \hspace{2mm}  \\ \hline
\hline
\end{tabular}
}
\caption{The values of mixing angles $\gamma_{\ell, d}$ for different types of 2HDM's. ``L" denotes the lepton-specific 2HDM (2HDM-L).  The ``flipped" 2HDM (2HDM-F) is similar to the 2HDM-II except that the charged leptons couple to the same Higgs as the up-type quarks.  
\label{tab:match}}
\end{table}
The Yukawa coupling values with respect to the SM can then be written generally as
\bea
\kv&=&\sin(\beta-\alpha),\\
\ku&=&\cos \alpha /\sin\beta,\\
\kd&=&-\sin(\alpha-\gamma_d)/\cos(\beta-\gamma_d),\\
\kl&=&-\sin(\alpha-\gamma_\ell)/\cos(\beta-\gamma_\ell),
\eea
where $\tan \beta = {v_2/ v_1}$ is the ratio of the doublet vacuum expectation values.   Specific values for the traditional 2HDMs are given in Table~\ref{tab:Yuk2HDM}.
\begin{table}[h]
\caption{Yukawa couplings of the fermions to the lightest Higgs with respect to the SM for traditional 2HDMs.}
\begin{center}
\begin{tabular}{|c|cccc|}
\hline
Model & $\kv$ & $\ku$ & $\kd$ & $\kl$\\
\hline
2HDM-I & $\sin (\beta-\alpha)$ & $\ca/\sb$ & $\ca/\sb$ & $\ca/\sb$ \\
2HDM-II & $\sin (\beta-\alpha)$ & $\ca/\sb$ & $-\sa/\cb$ & $-\sa/\cb$ \\
2HDM-L & $\sin (\beta-\alpha)$ & $\ca/\sb$ & $\ca/\sb$ & $-\sa/\cb$ \\
2HDM-F & $\sin (\beta-\alpha)$ & $\ca/\sb$ & $-\sa/\cb$ & $\ca/\sb$ \\
\hline
\end{tabular}
\end{center}
\label{tab:Yuk2HDM}
\end{table}%

We pause to note that while our parametrization is similar to that of~\cite{Bai:2012ex}, we have chosen here to work in a different basis because it facilitates the comparison to the conventions adopted in much of the previous 2HDM work, including~\cite{Branco:2011iw}.   More precisely, it is well known that one can use a $U(2)$ transformation to select a specific basis for the two linearly independent Higgs states (see e.g.~\cite{Davidson:2005cw} for a comprehensive discussion).  The basis chosen in \cite{Bai:2012ex} is one in which the mixing angle for the lepton couplings was set to zero, such that both up-type and down-type quarks couple to both Higgs doublets with nontrivial mixing angles $\gamma_u'$ and $\gamma_d'$, respectively (note that in this basis, the Type I model necessarily has all fields coupling to $\Phi_1$ rather than $\Phi_2$, and the flipped model also has $\Phi_1$ and $\Phi_2$ interchanged).  Our basis is related to the basis $\Phi_{1,2}'$ of \cite{Bai:2012ex} via the transformation $\Phi_{1,2}= \cos \gamma_{\ell} \Phi_{1,2}'-\sin \gamma_{\ell} \tilde{\Phi}_{2,1}'$.   Note that the basis-dependent parameter $\tan\beta$ differs for the two cases: our $\tan\beta$ is related to the $\tan\beta'$ of \cite{Bai:2012ex} via $\tan\beta=\tan(\beta'-\gamma_\ell)$.  Of course, the physics is independent of any specific basis choice.

\begin{figure}[!t]
\begin{center}
     \includegraphics[angle=0,width=0.47\textwidth]{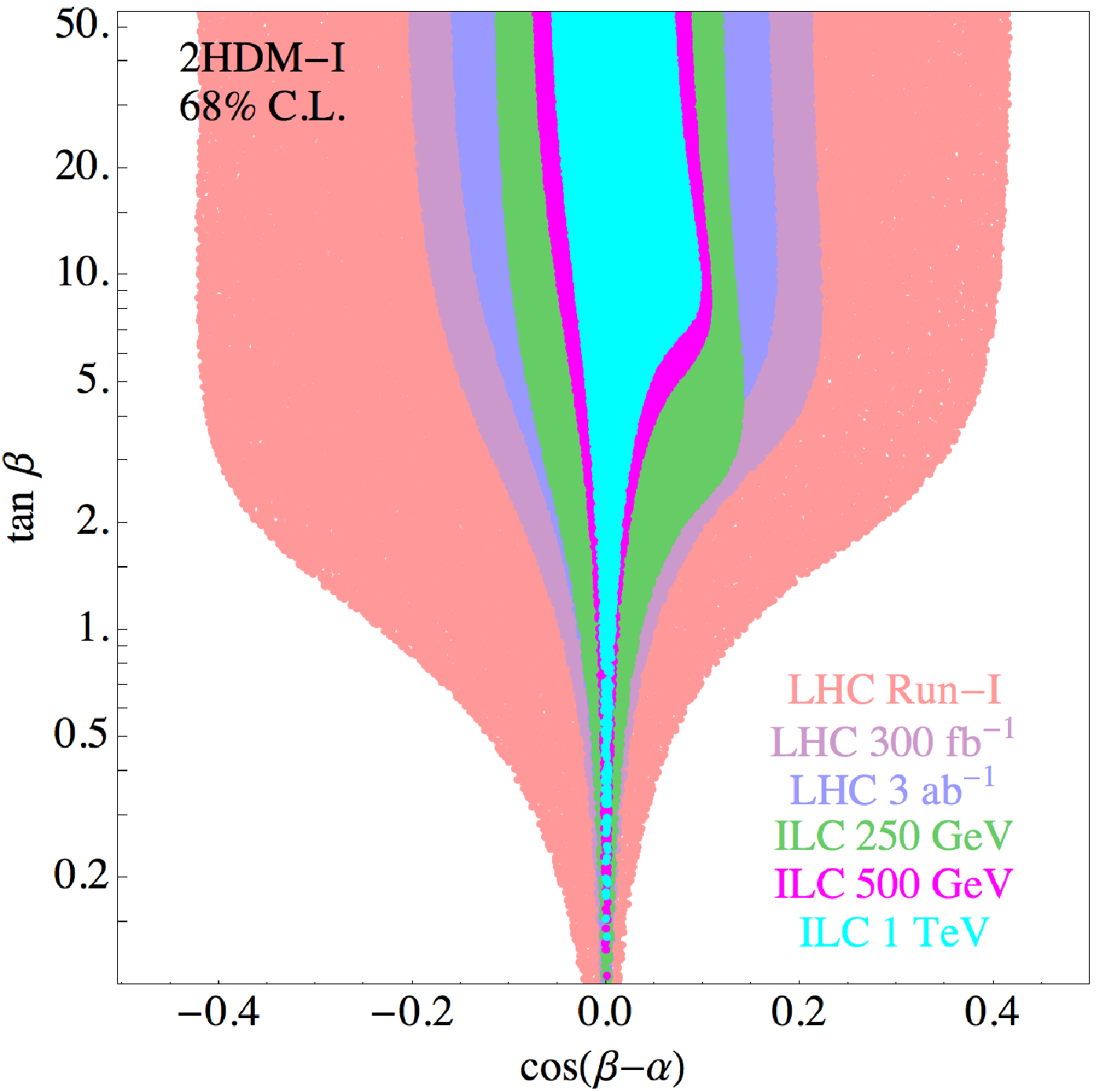}
     \includegraphics[angle=0,width=0.47\textwidth]{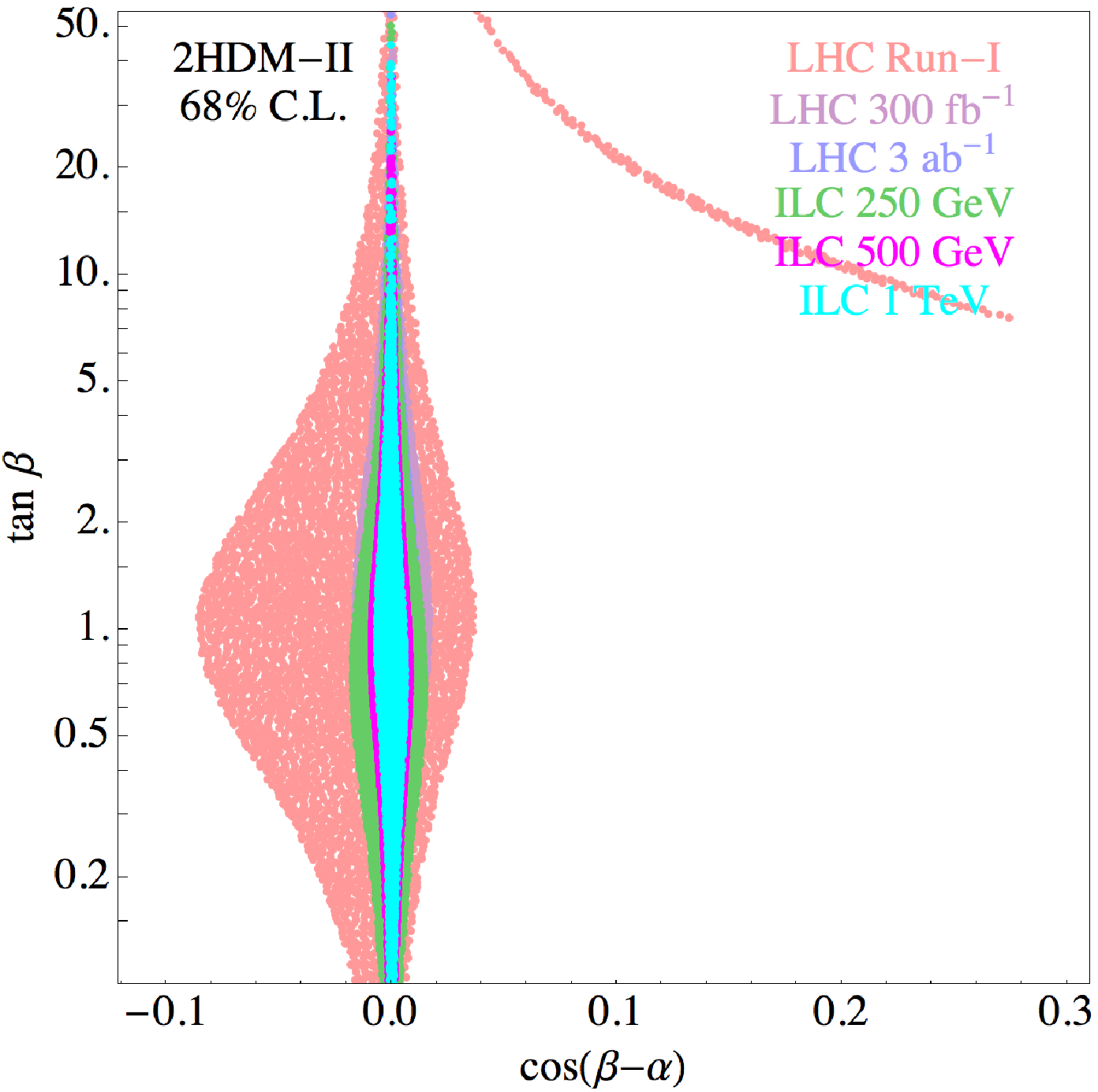}\\
     \includegraphics[angle=0,width=0.47\textwidth]{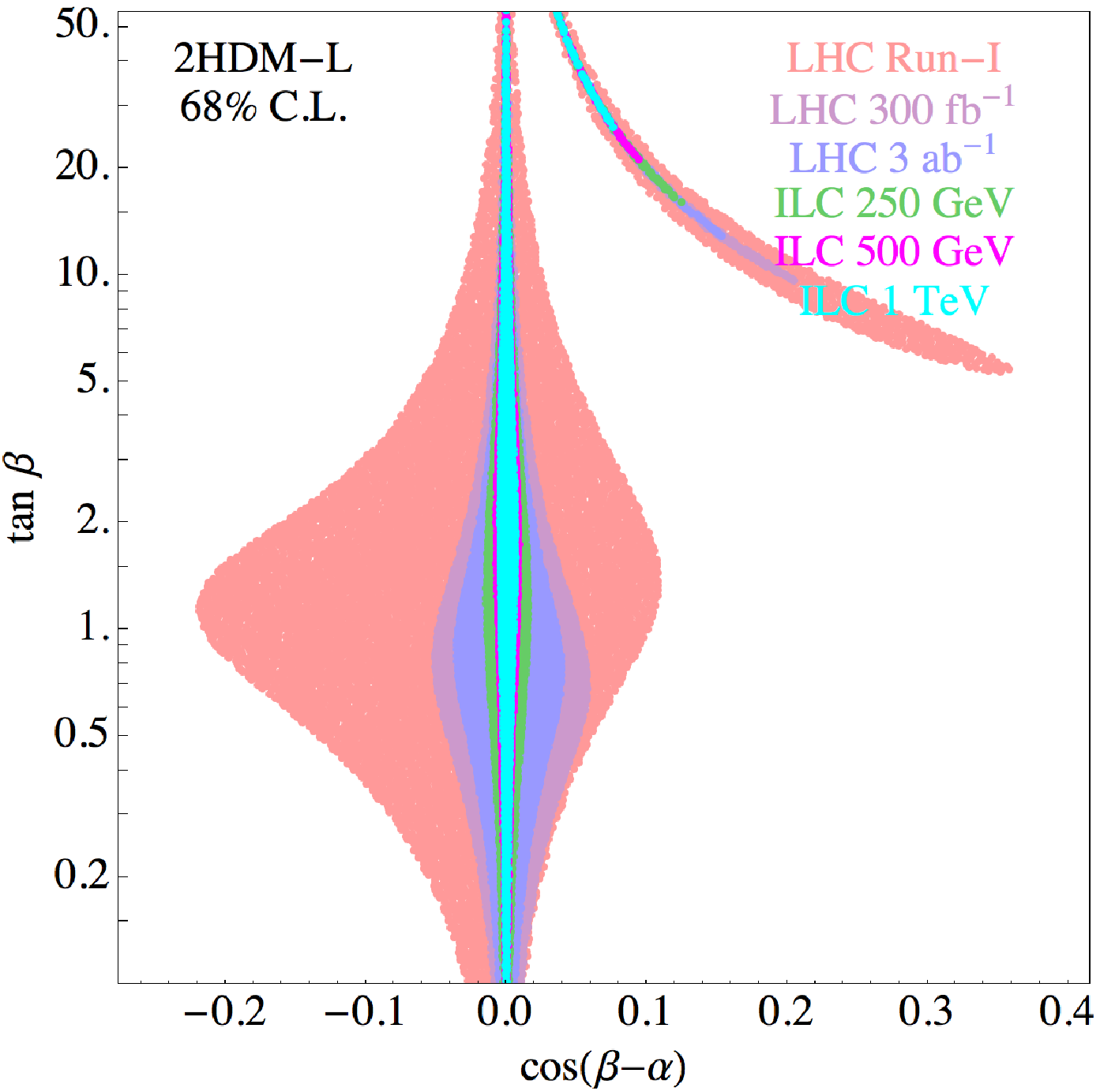}
     \includegraphics[angle=0,width=0.47\textwidth]{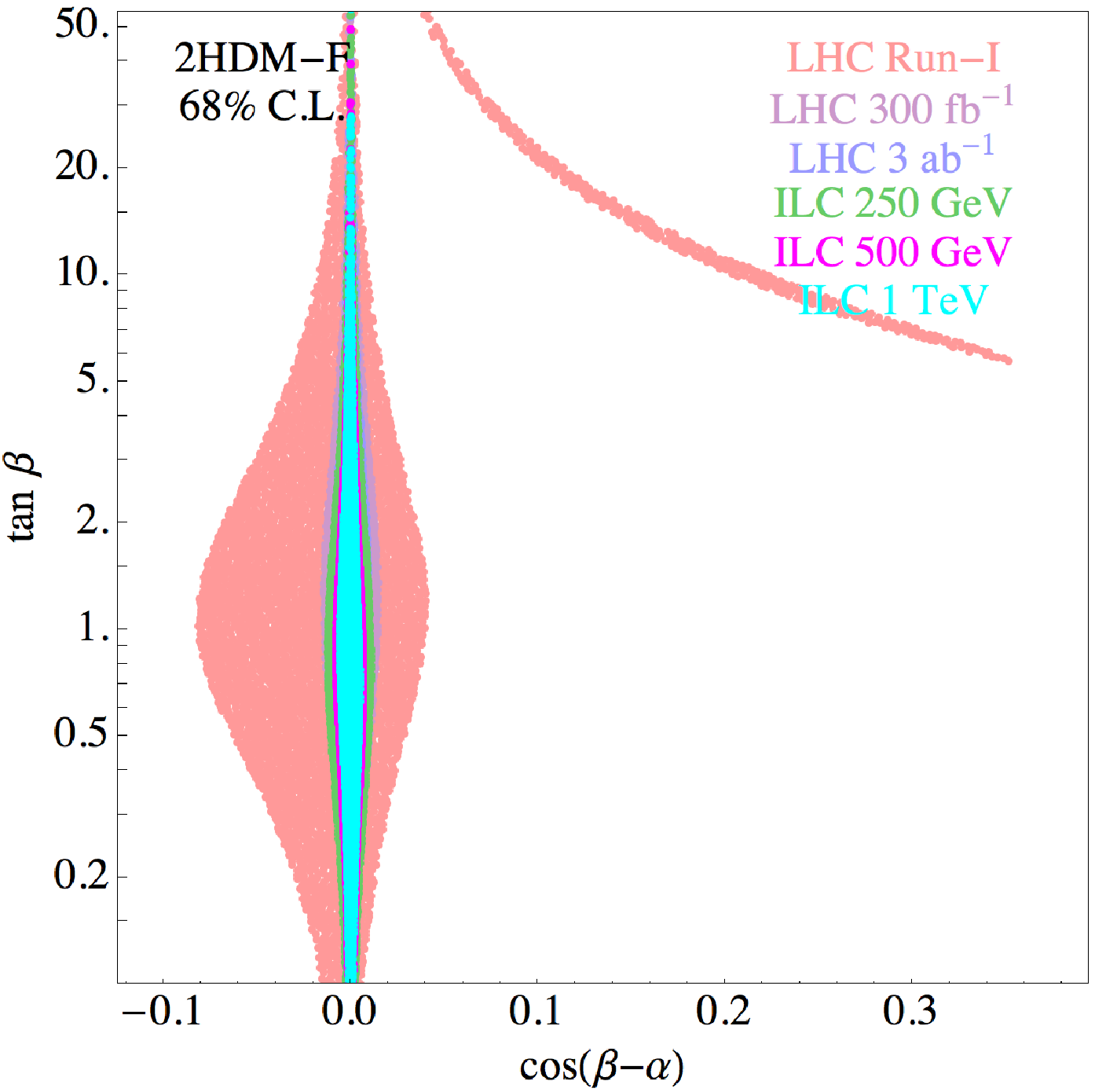}
\caption{The current $\tan\beta$ vs. $\cos(\beta-\alpha)$ regions inhabited at the 1$\sigma$ level for the usual two Higgs doublet models: a) 2HDM-I, b) 2HDM-II, c) 2HDM-L, d) 2HDM-F.}
\label{fig:delta_tb68}
\end{center}
\end{figure}
\begin{figure}[!t]
\begin{center}
     \includegraphics[angle=0,width=0.47\textwidth]{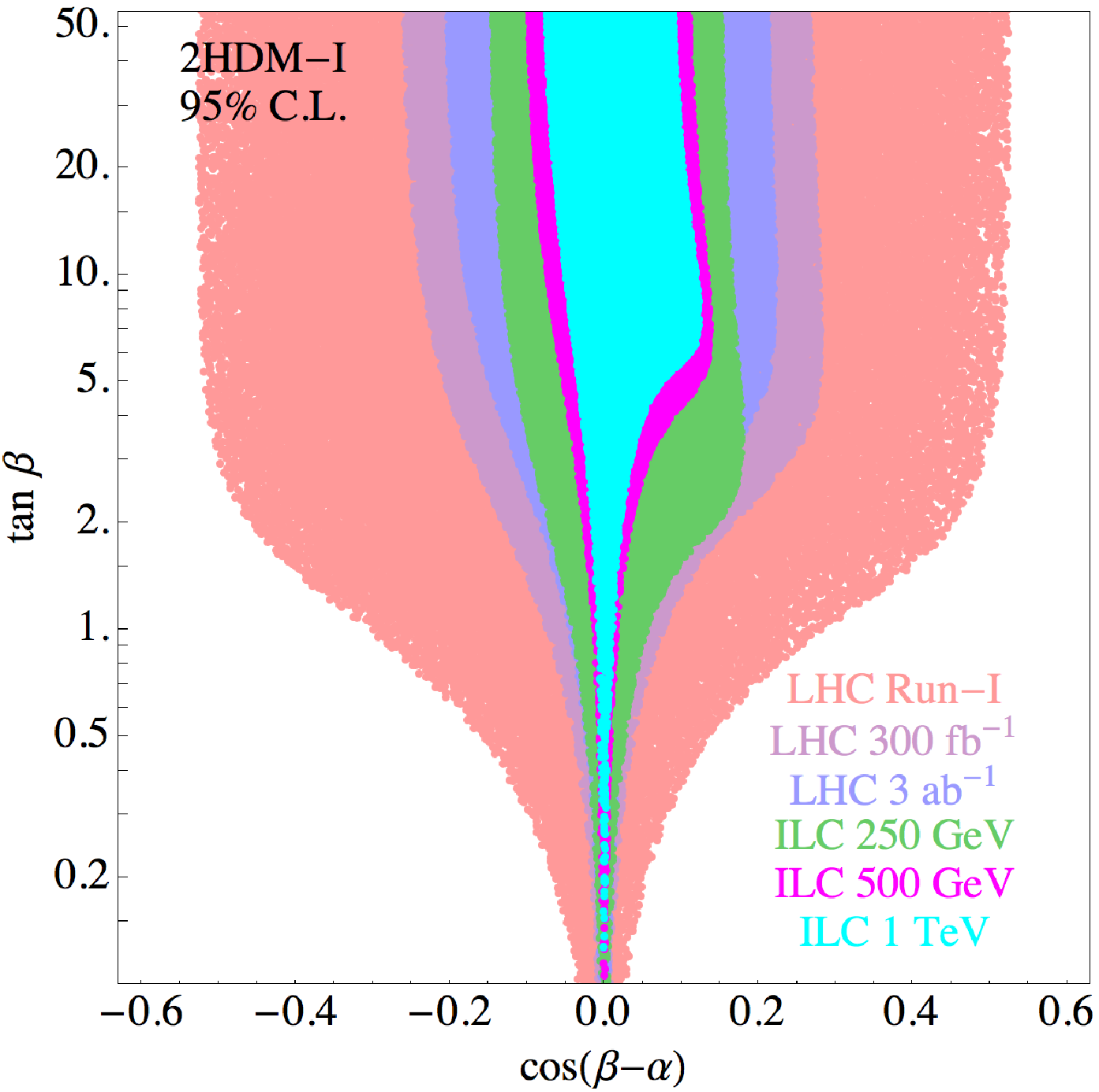}
     \includegraphics[angle=0,width=0.47\textwidth]{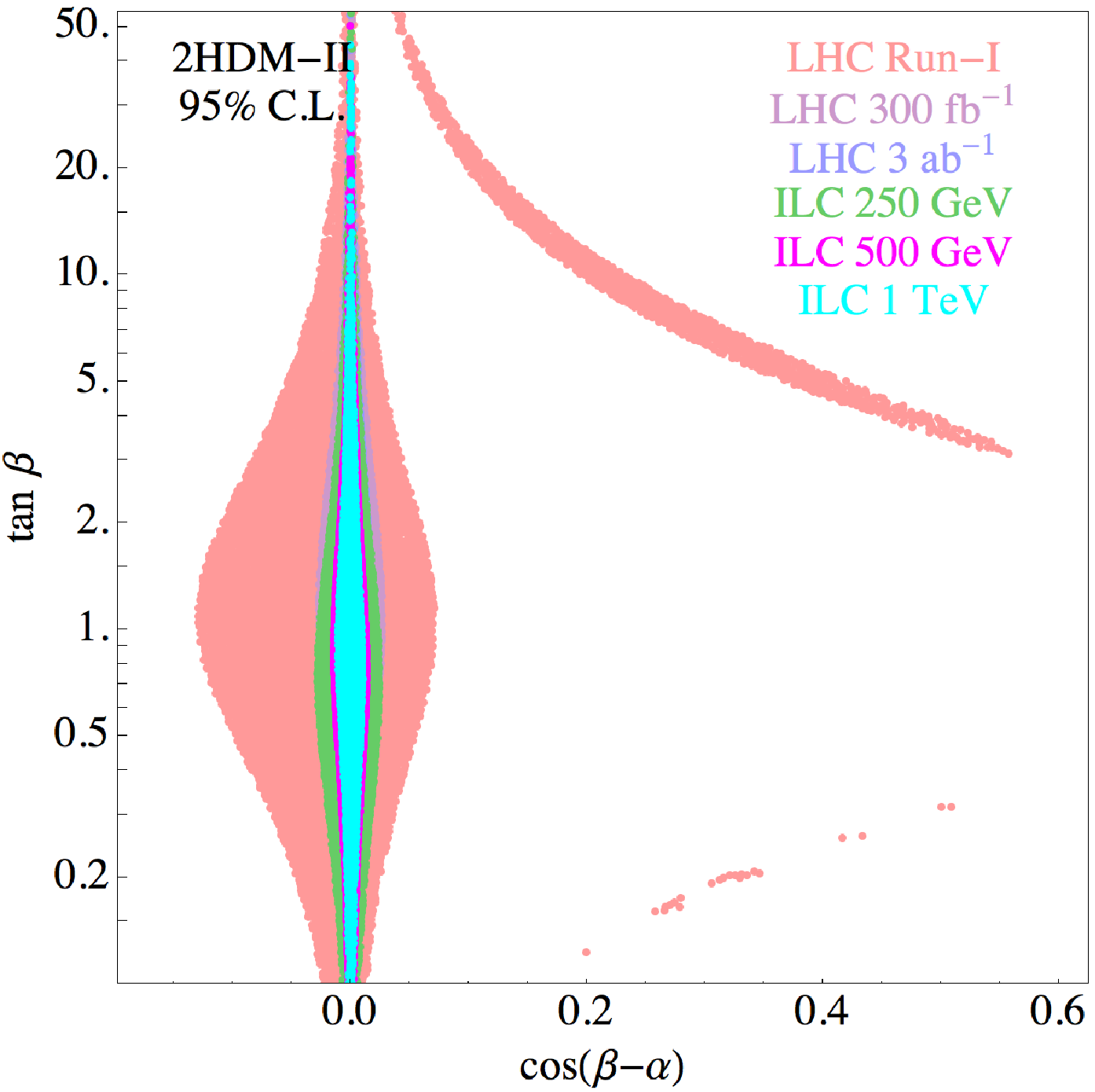}\\
     \includegraphics[angle=0,width=0.47\textwidth]{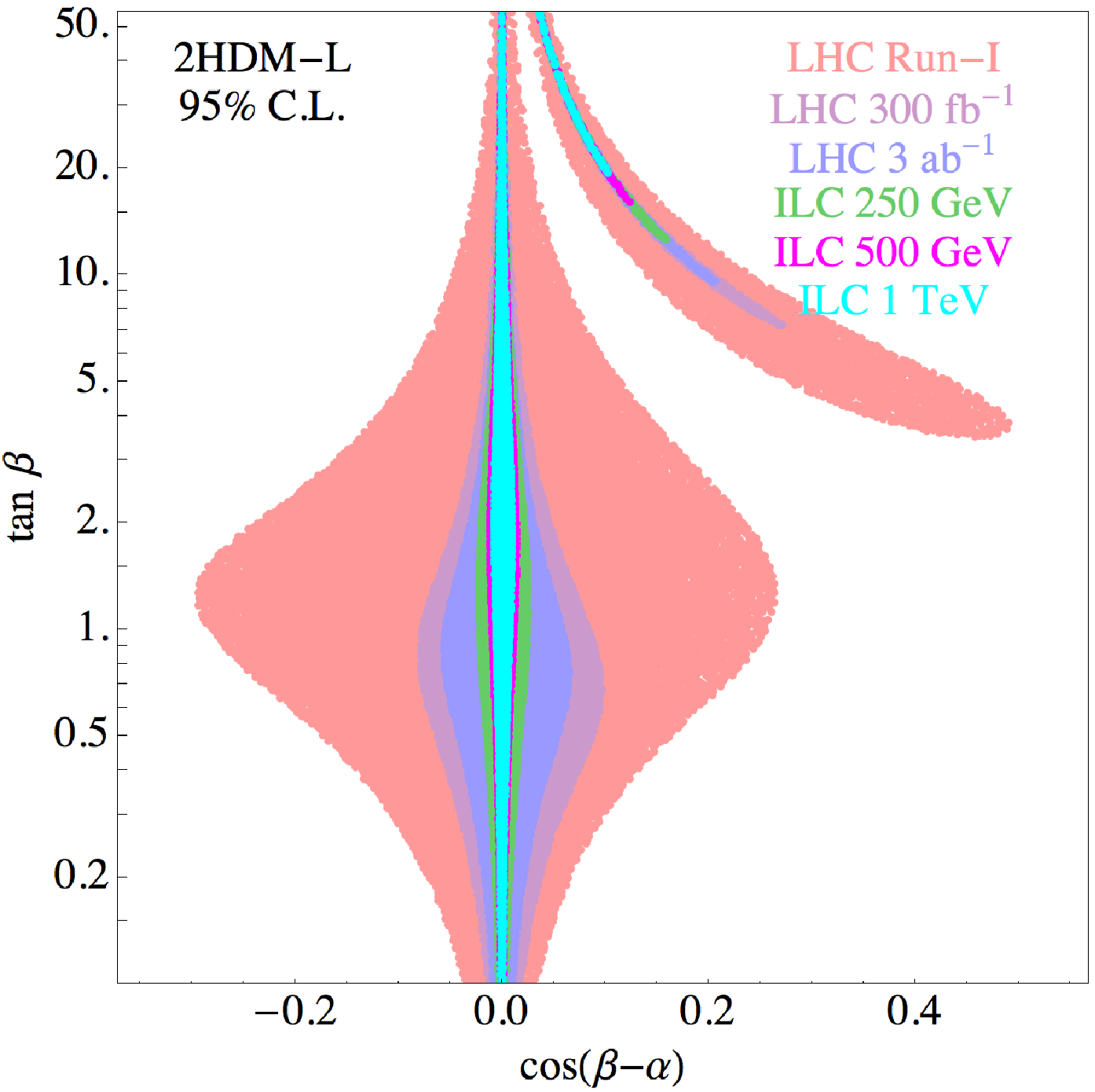}
     \includegraphics[angle=0,width=0.47\textwidth]{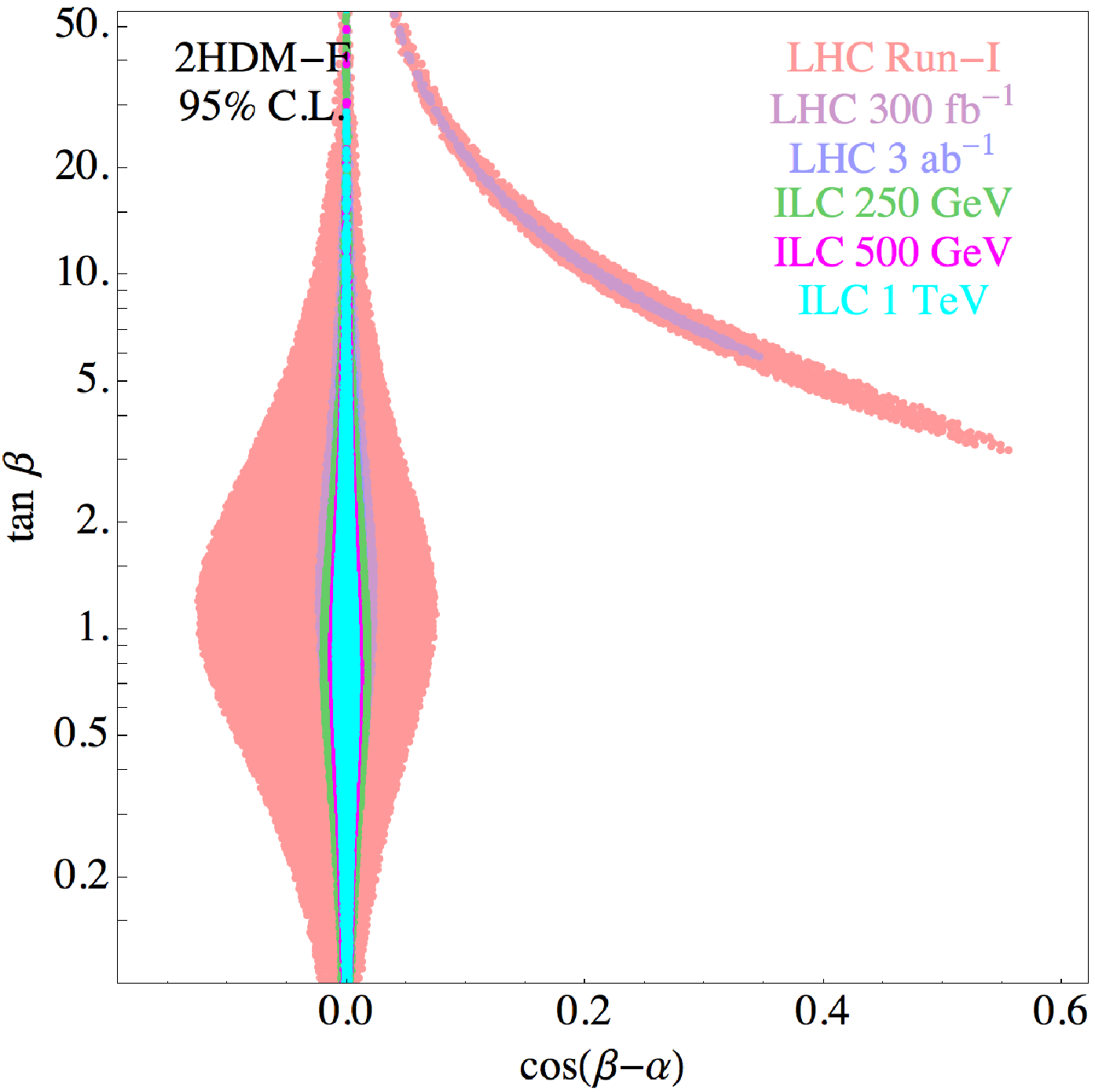}
\caption{Same as Fig.~\ref{fig:delta_tb68}, but at the 95\% C.L.  }
\label{fig:delta_tb95}
\end{center}
\end{figure}

\begin{figure}[!t]
\begin{center}
     \includegraphics[angle=0,width=0.47\textwidth]{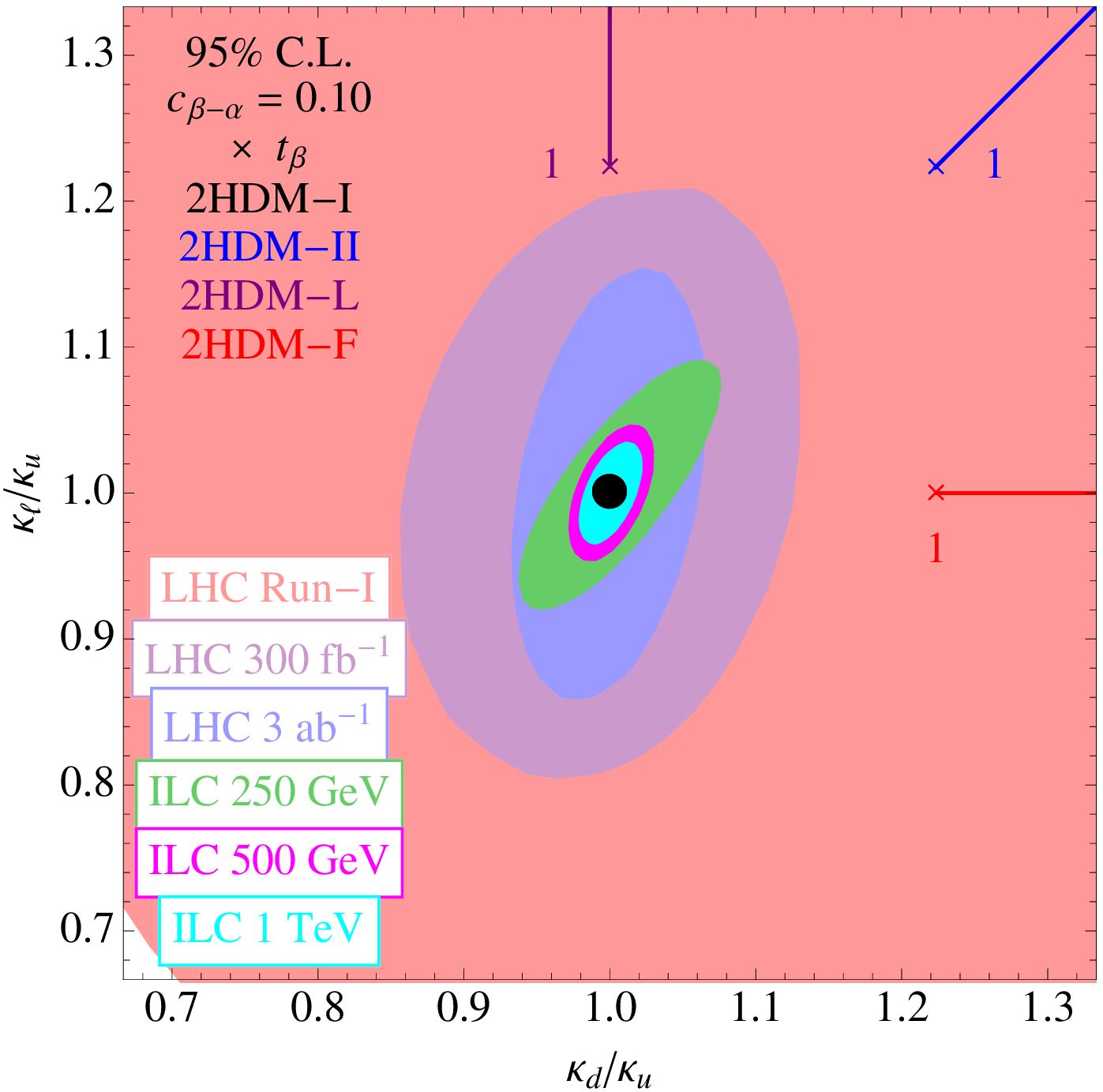}
     \includegraphics[angle=0,width=0.47\textwidth]{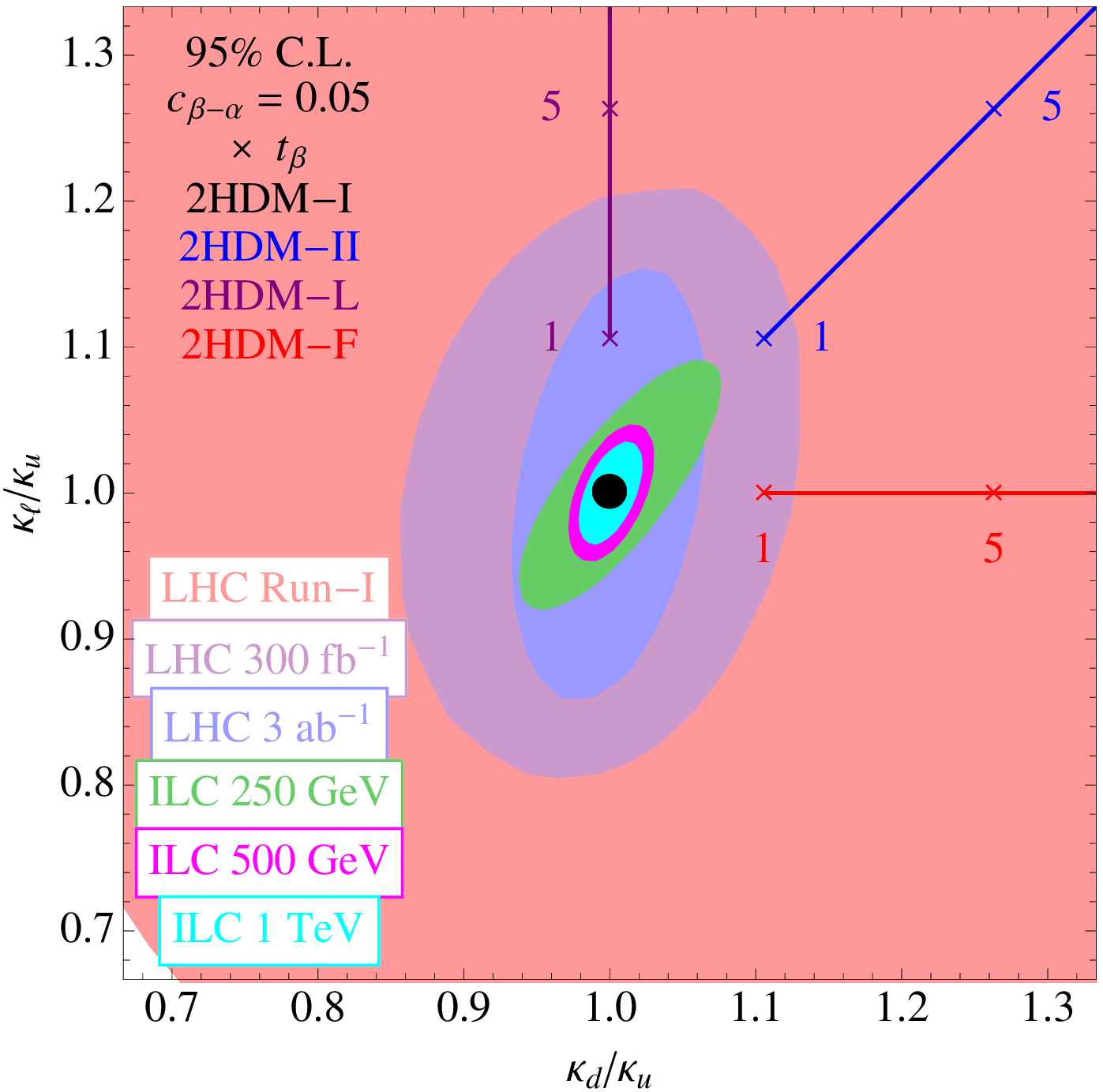}\\
     \includegraphics[angle=0,width=0.47\textwidth]{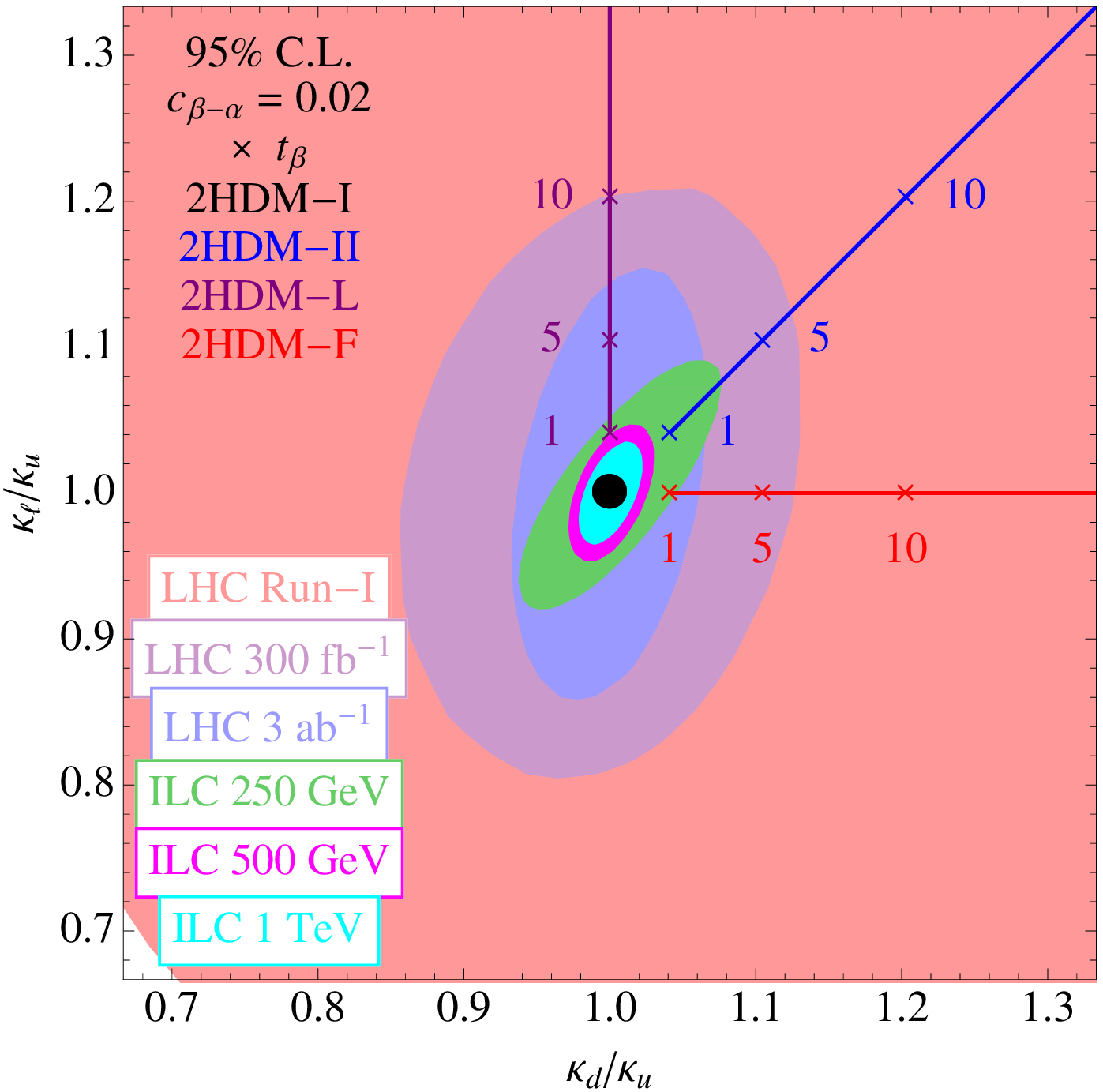}
     \includegraphics[angle=0,width=0.47\textwidth]{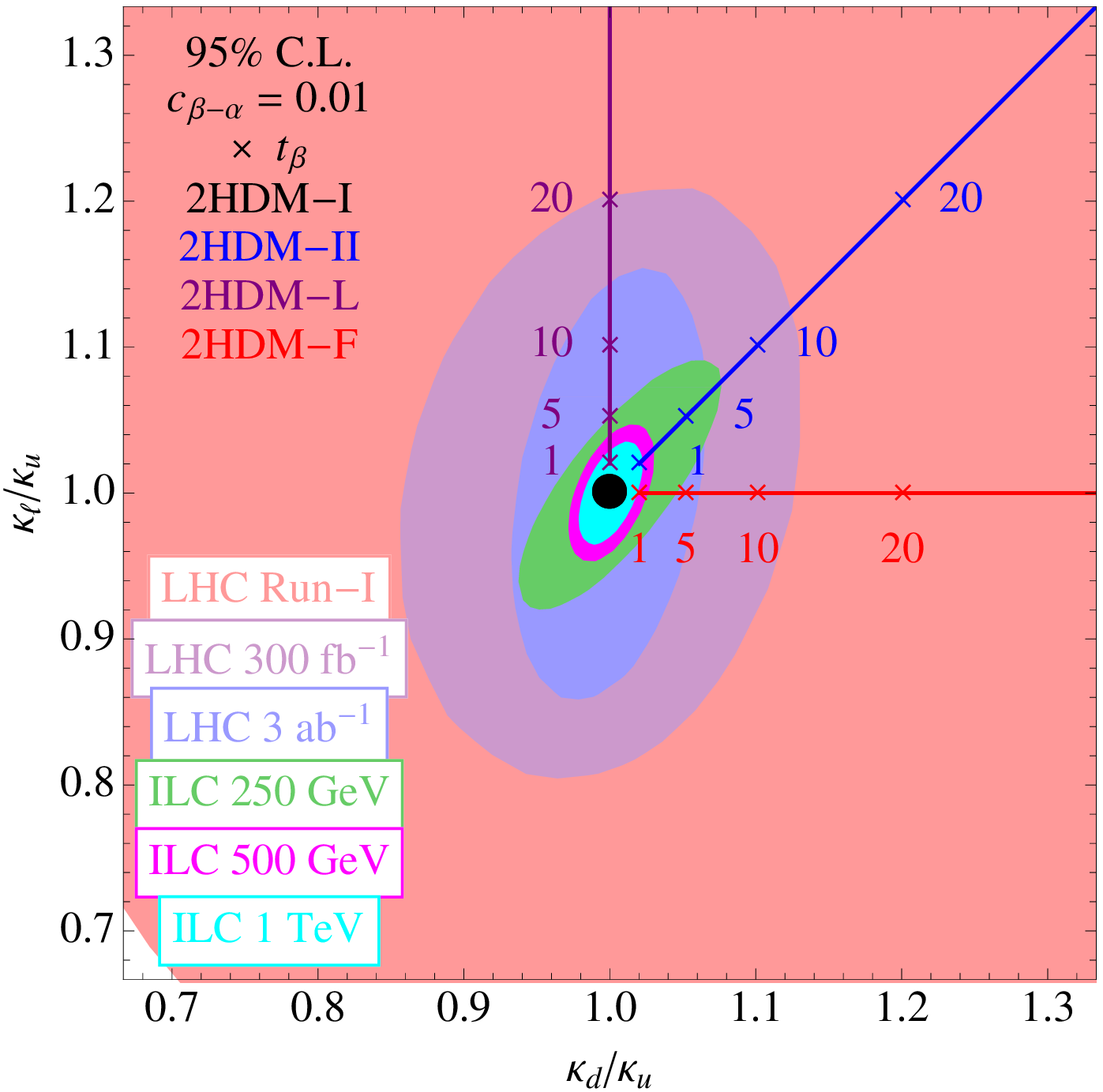}
\caption{Yukawa coupling ratios at the 95\% C.L. from a fit to current collider data and future LHC14 and ILC data.  The allowed regions are compared with the 2HDM predictions in terms of the decoupling parameter $\delta$. Filled regions match the notation of Fig~\ref{fig:genfitdiff}.}
\label{fig:ratio_delta}
\end{center}
\end{figure}

\begin{figure}[!t]
\begin{center}
     \includegraphics[angle=0,width=0.47\textwidth]{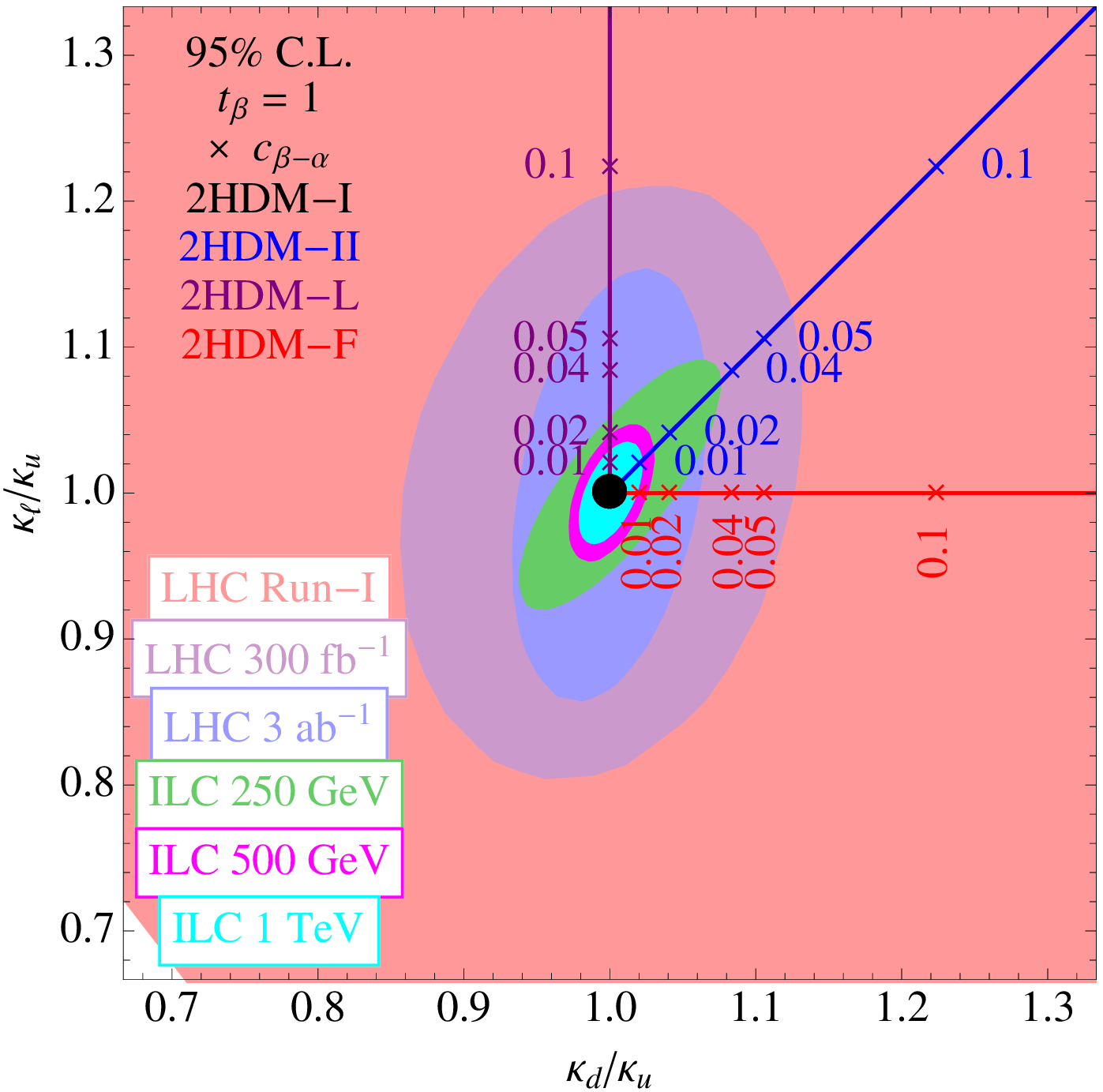}
     \includegraphics[angle=0,width=0.47\textwidth]{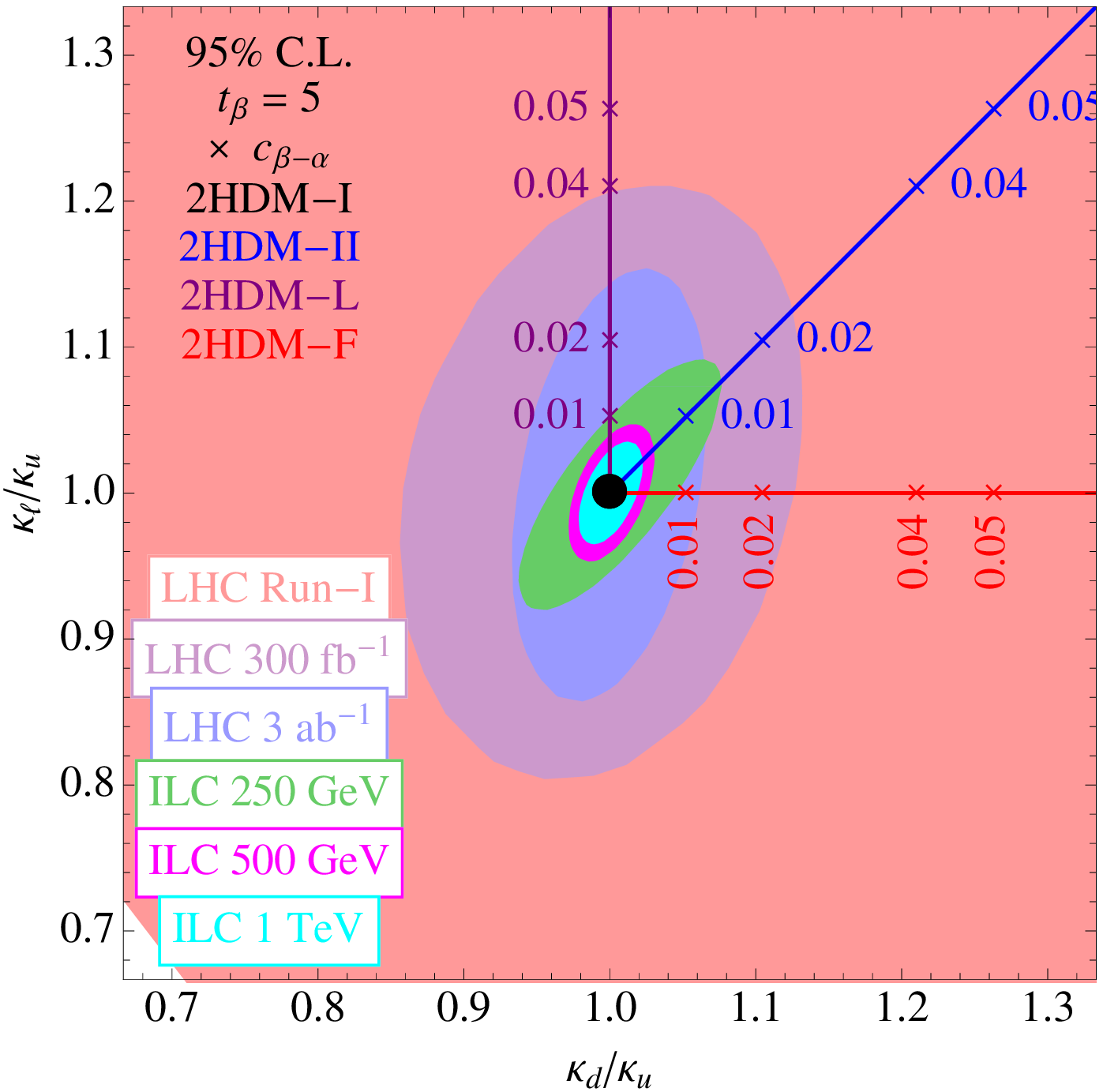}\\
     \includegraphics[angle=0,width=0.47\textwidth]{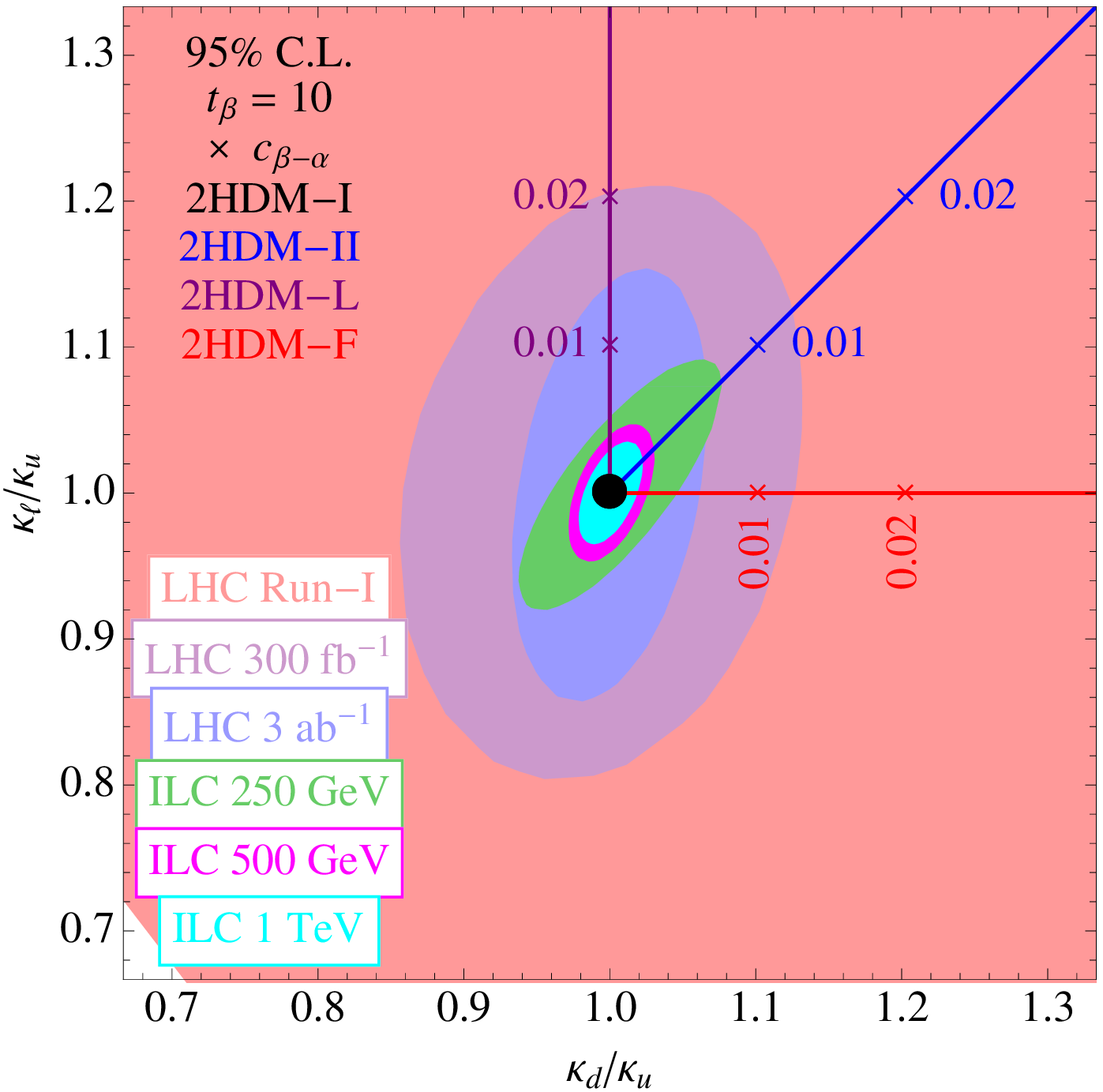}
     \includegraphics[angle=0,width=0.47\textwidth]{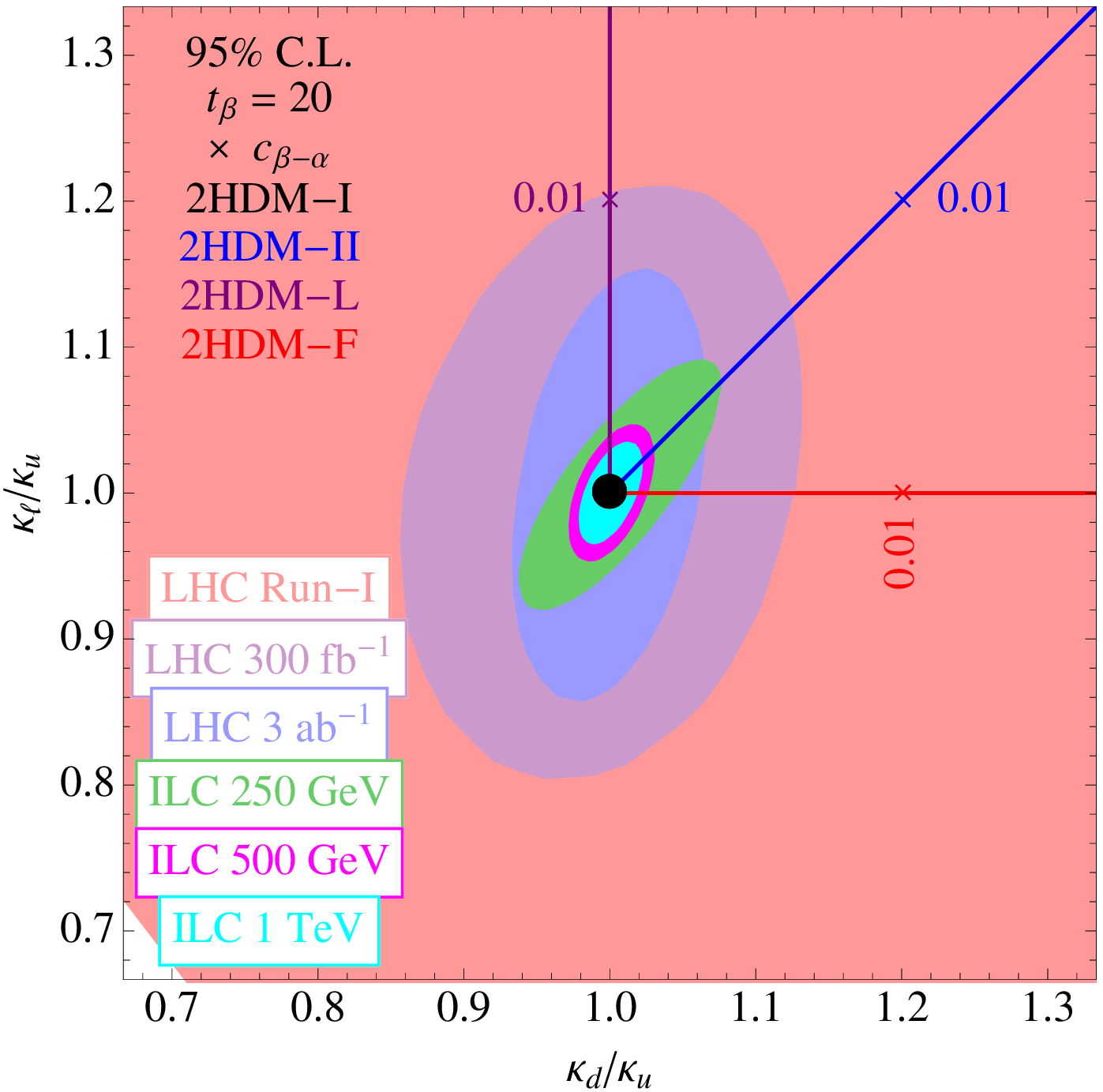}
\caption{Yukawa coupling ratios at the 95\% C.L.  from current data and future LHC14 and ILC data with the expected 2HDM predictions at given values of $\tan \beta$. The black dot refers to the 2HDM-I model, which in this plane matches the SM.  The 95\% C.L. filled regions match those of the general fit in Fig.~\ref{fig:genfitdiff}b.}
\label{fig:ratio_tb}
\end{center}
\end{figure}
We now trade the mixing angle $\alpha$ for the decoupling parameter $\delta\equiv\cos(\beta-\alpha)$, which parameterizes the coupling of the heavy CP-even Higgs to electroweak Vector Bosons as
\be
\kv = \sin (\beta-\alpha) = \sqrt{1-\delta^2}.
\ee
This replacement allows us to write the Yukawa couplings as
\bea
\ku&=&\sin(\beta- \theta_V) / \sin(\beta)\\
\kd&=&\cos(\beta-\gamma_d- \theta_V) / \cos(\beta-\gamma_d)\\
\kl&=&\cos(\beta-\gamma_\ell- \theta_V) / \cos(\beta-\gamma_\ell)
\eea
where $\theta_V = \sin^{-1} \delta$; in the limit of $\delta \to 0$, the SM Higgs doublet coupling is recovered.  For the traditional models, we show in Fig.~\ref{fig:delta_tb68} the 1$\sigma$ regions that are consistent with current LHC data along with  projected sensitivities at  LHC14 with 300 fb$^{-1}$ of integrated luminosity and with the future ILC scenarios.  The 95\% C.L. regions are presented in Fig.~\ref{fig:delta_tb95}.  The current LHC exclusion regions and the projections to 300 and 3000 \ifb agree with those of Ref.~\cite{Chen:2013rba}.  The curved branches extending from the decoupling region with $\cos(\beta-\alpha) > 0$ correspond to points where $\kd$ and/or $\kl$ flip sign.  This affects the cross sections measurable at the LHC in a small way, and may probed at the ILC.

Distinguishing various flavors of 2HDMs can be done via the Yukawa coupling ratios.  For example, using the above relations, we find the coupling ratios that deviate from the SM do so in the following manner for 2HDM-II near the decoupling limit, $\delta \to 0$:
\be
{\kappa_{\ell,d} \over \ku} = {-\cot\alpha\over \cot\beta} = 1 + {-2\delta\over \sin2\beta}   + \delta^2\left(1+\tan^2\beta\right)  + {\cal O}\left(\delta^3\right).
\label{eq:ratiodecoup}
\ee

These limits are shown in their exact form, for each model, in Fig.~\ref{fig:ratio_delta} for fixed $\tan\beta$ and in Fig.~\ref{fig:ratio_tb} for fixed $\delta$.  Any model parameter deviations from those shown in these figures can be extracted using Eq.~\ref{eq:ratiodecoup}.  The model expectations are overlain on the 95\% C.L. regions from the general Yukawa coupling fit described in Section~\ref{sect:genfit}.  

\begin{figure}[!b]
\begin{center}
     \includegraphics[angle=0,width=0.7\textwidth]{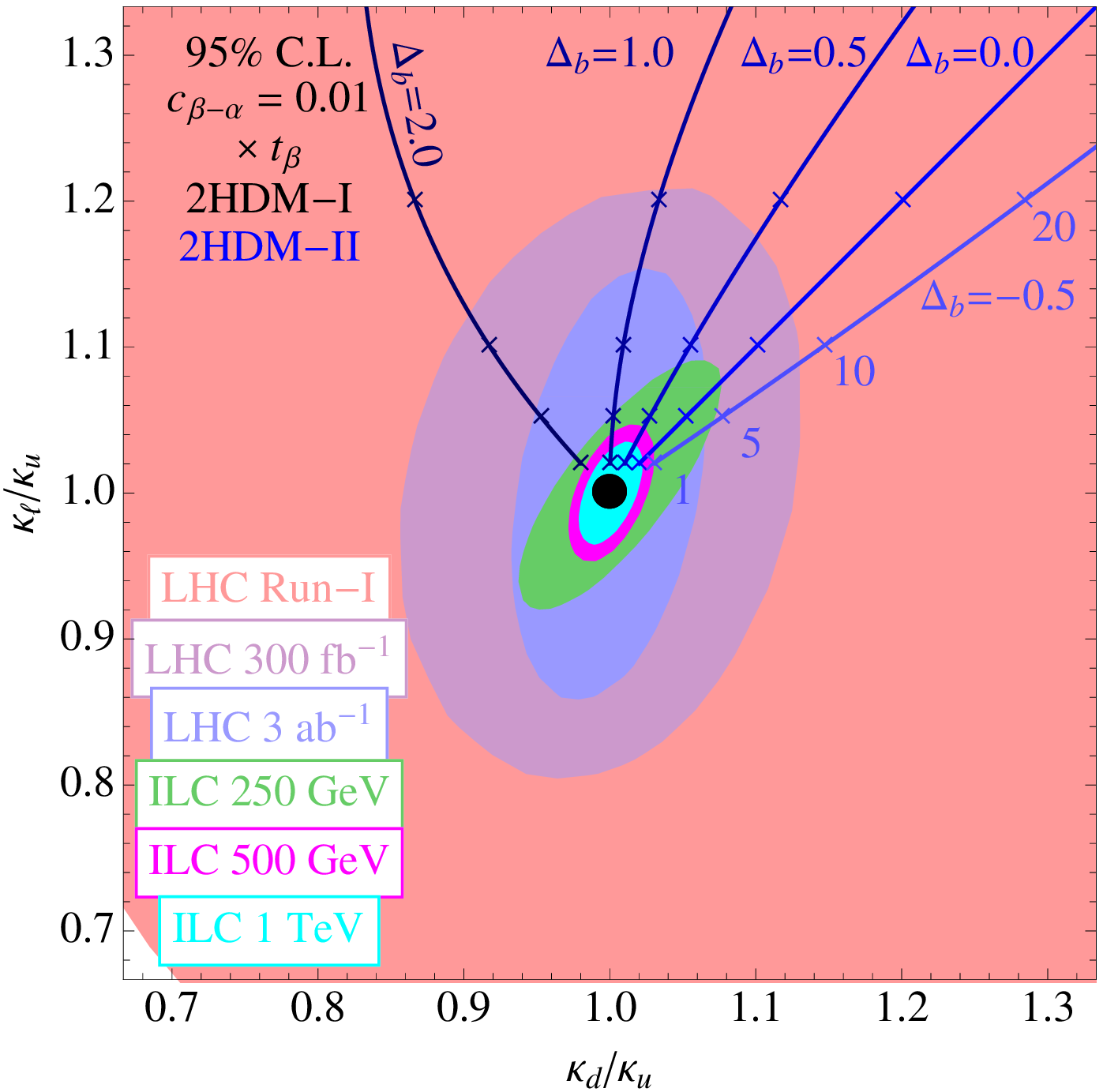}
\caption{Yukawa coupling ratios at the 95\% C.L.  to current data and future LHC14 and ILC data with the expected predictions within the 2HDM-II with a loop induced ``wrong Higgs'' coupling, $\Delta_b$.  The 95\% C.L. filled regions match those of the general fit in Fig.~\ref{fig:genfitdiff}b.  The spacing of the $\Delta_b$ curves scale with $\cos(\beta-\alpha)$.}
\label{fig:delb}
\end{center}
\end{figure}
The coupling of a fermion to the ``wrong Higgs'' may occur in some models.  For instance, in the MSSM, the loop induced coupling of the up Higgs doublet to the bottom quark can be induced via loops of gluinos, squarks and higgsinos~\cite{Carena:1994bv,Carena:2001bg,Barger:2012ky,Hagiwara:2012mga,Carena:2013iba}.  The bottom quark mass is shifted according to
\be
m_b = {y_b v_1 + \Delta y_b v_2\over \sqrt 2} = {y_b v \cos\beta\over\sqrt 2}\left(1+\Delta_b\right).
\ee
In the presence of these corrections, the bottom Yukawa coupling becomes
\be
\kd \to \kd +(1+ \cot\alpha \cot\beta) \Delta_b.
\ee
Note that in the decoupling limit, the additional contributions vanish (c.f. Eq.~\ref{eq:ratiodecoup}).  Given the present exclusion limits from the LHC on sparticle masses, the value of this wrong Higgs coupling is suppressed, but ${\cal O}(1)$ corrections are still possible. Indeed, one can extract the shift in the bottom coupling from,
\be
\Delta_b \approx {\kd - \kl \over \ku-\kd}.
\ee
Observe that the correlation in the Yukawa coupling ratios is strongest in the $\kd/\ku\approx \kl/\ku$ direction.  As a result, there is sensitivity to deviations off  the 2HDM-II line shown in Fig.~\ref{fig:delb}, indicating a sensitivity to such a loop-induced wrong Higgs coupling.

\section{Width Measurements}
\label{sect:width}

The Higgs boson width can be measured indirectly at the LHC via comparison of its relative production and decay rates.  However, this assumes that the Vector Boson coupling cannot be much larger than that given by the SM, typically $\kv < 1.05$~\cite{Duhrssen:2004cv,Barger:2012hv}.  This is motivated by two arguments: (i) Unitarity within doublet models requires that $\kv \leq 1$.  The limit is relaxed in fits to allow for measurement uncertainties.  (ii) Without an absolute cross section measurement, there exists a flat direction in general fits if one allows the total width to float.  At the ILC, this coupling can be directly determined from the $Z$-Higgsstrahlung cross section measurement.  Therefore, the Higgs width can be inferred via comparison of the measured production cross sections relative to that of $\sigma_{Zh}$.  

\begin{figure}[htbp]
\begin{center}
     \includegraphics[angle=0,width=0.47\textwidth]{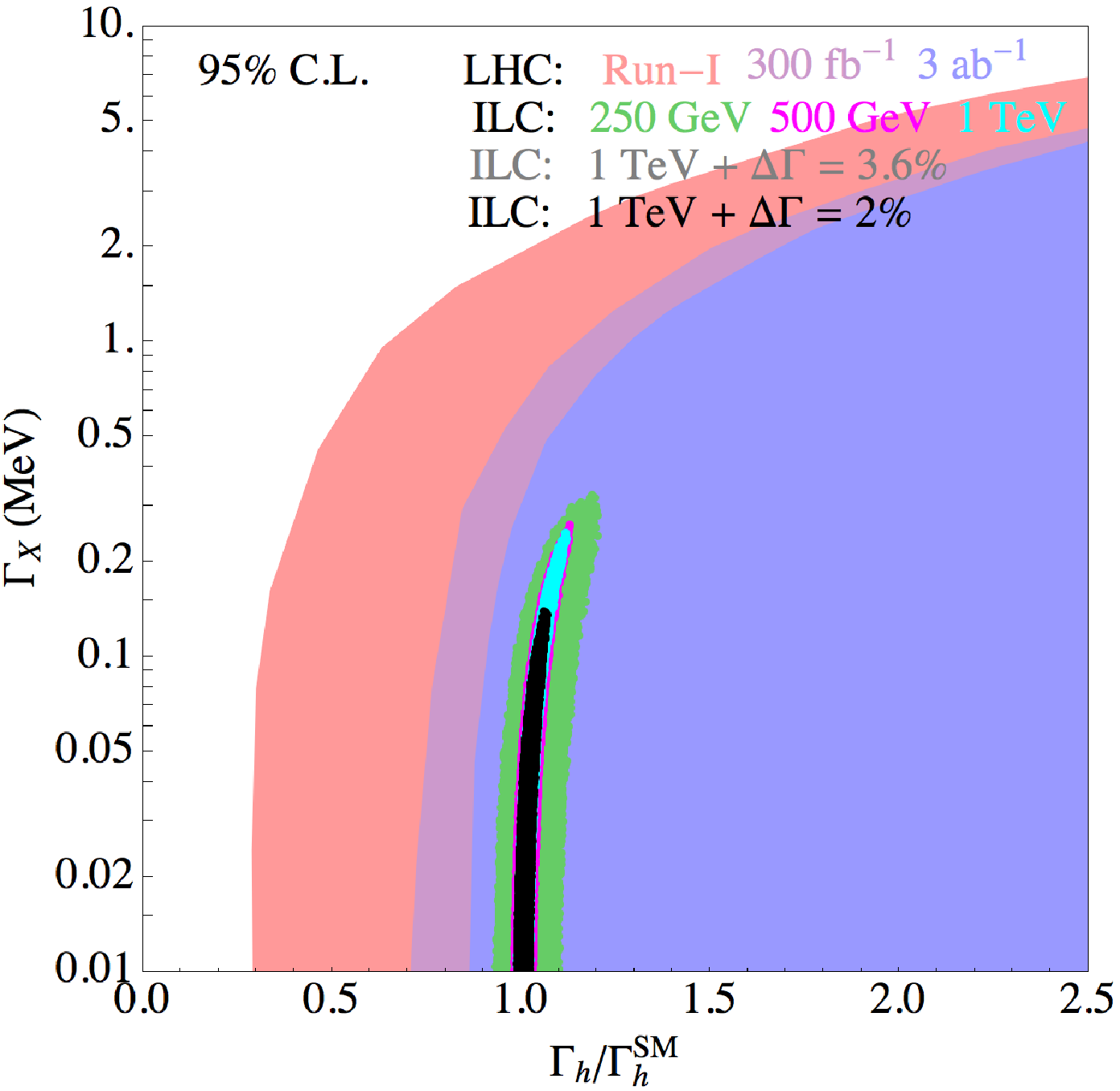}
     \includegraphics[angle=0,width=0.47\textwidth]{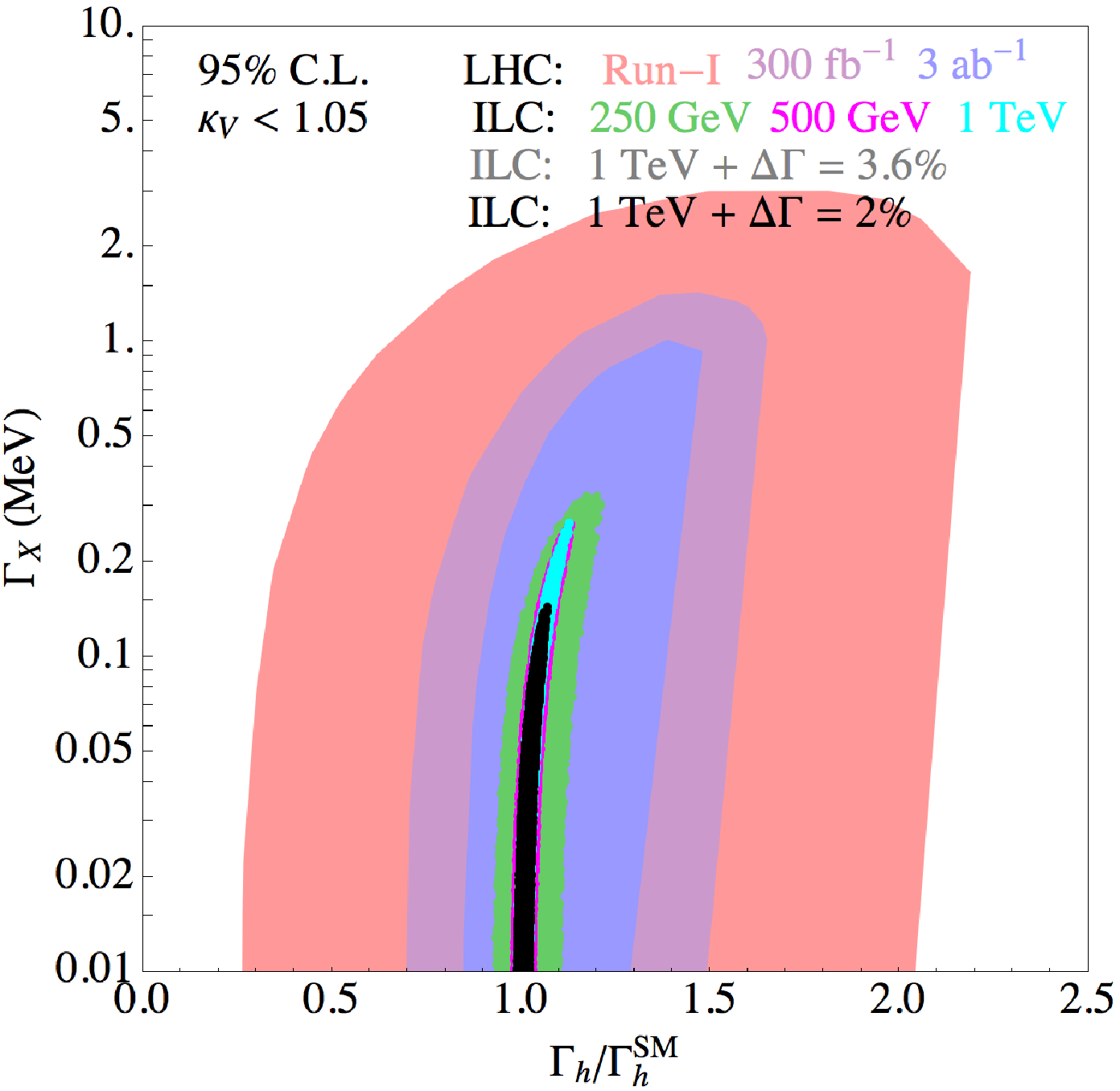}\\
     \includegraphics[angle=0,width=0.47\textwidth]{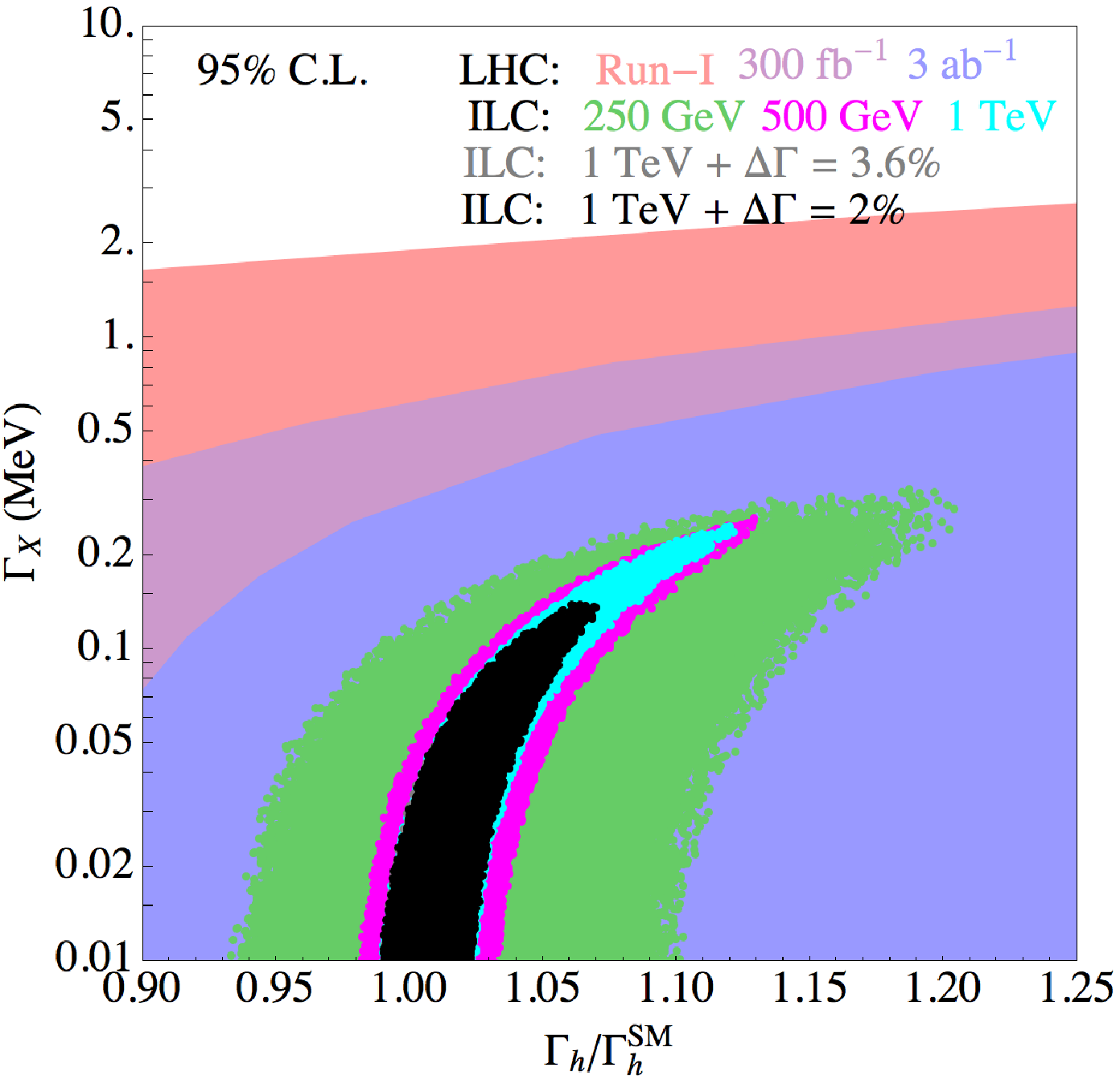}
\caption{Regions of the total width, $\Gamma_h^{\rm meas}/\Gamma_h^{\rm SM}$, and the additional partial width into non-SM modes, $\Gamma_X$, at the 95\% C.L.  We assume (a) $\kv$ free, (b) $\kv < 1.05$.  Panel (c) is the same as (a), but with emphasis on the ILC.  }
\label{fig:width}
\end{center}
\end{figure}
We perform two fits that allow parameters $\kv, \ku, \kd$, and $\kl$ described in Section~\ref{sect:genfit}, along with an additional new physics decay width, $\Gamma_X$, to vary.  The total width, $\Gamma_h$ is computed from the SM partial  and new physics widths as follows
\be
\Gamma_h = \kv^2 \left(\Gamma^{\rm SM}_{W^+W^-}+\Gamma^{\rm SM}_{ZZ}\right) + \kd^2  \Gamma^{\rm SM}_{b\bar b}+\kl^2 \Gamma^{\rm SM}_{\tau^+\tau^-} + \ku^2 \Gamma^{\rm SM}_{c\bar c} + \kg^2 \Gamma^{\rm SM}_{gg}+\ka^2 \Gamma^{\rm SM}_{\gamma\gamma} + \Gamma_X.
\ee
One fit requires $\kv < 1.05$, while the other is without this restriction.  In Fig.~\ref{fig:width}, we show the 2d regions of $\Gamma_h$, scaled by the SM value, and $\Gamma_X$ for various machine assumptions at the 95\% C.L.  Similar results of the 1d $\Delta \chi^2$ fit are presented in Table~\ref{tab:width}.
\begin{table}[htdp]
\caption{Width measurement uncertainties at the $1\sigma$ from a 1d fit over $\kv, \ku, \kd, \kl$ and $\Gamma_X$, the new physics decay width, for the collider benchmarks assuming the absence of non-standard decay modes. We assume $\kv < 1.05$ for only the LHC, and lift this restriction for the ILC and MC benchmarks. }
\begin{center}
\begin{tabular}{|c|cc|}
\hline
 & $\delta \Gamma_h/\Gamma_h^{\rm SM}$ & $\Gamma_X$  (MeV)  \\
 \hline
 	LHC-I   & $^{+0.30}_{-0.44}$&  $ < 2.3$\\
	LHC300  		& $^{+0.36}_{-0.16}$ & $<1.2$ \\
	LHC3000  & $^{+0.30}_{-0.08}$& $<0.8$\\
\hline
	ILC250  & $^{+0.075}_{-0.037}$ & $<0.25$\\
	ILC500  & $^{+0.052}_{-0.012}$ & $<0.21$\\
	ILC1000  & $^{+0.051}_{-0.008}$ & $<0.21$\\
\hline
	+ MC 3.6\% &  $^{+0.026}_{-0.009}$ & $<0.11$\\
	+ MC 2.0\% &  $^{+0.016}_{-0.008}$ & $<0.07$\\
\hline
\end{tabular}
\end{center}
\label{tab:width}
\end{table}%

At a future muon collider (MC), a scan over the line shape of the Higgs resonance will quickly discover the Higgs boson.  The line scan can also directly determine the total Higgs width.  Depending on the luminosity delivered in each step of the resonance scan, the MC is anticipated to have a 3.6\% uncertainty with the default configuration, $\int{\cal L} dt = 0.05$ \ifb~per step with 20 steps over a 60 MeV energy range, assuming a beam energy resolution of 0.003\%.  The uncertainty shrinks to 2\%  with a four-fold increase in luminosity per step~\cite{Barger:1995hr,Han:2012rb,MCHiggs}.   Similar scans for the associated heavy states can provide a wealth of information~\cite{Eichten:2013ckl}.

\section{Testing for additional doublets and singlets}
\label{sect:pattern}

The so-called pattern relations of the measured couplings scaled with respect to the SM are defined by~\cite{Ginzburg:2001wj,Ginzburg:2001ss,Barger:2009me}
\bea
P_{ul} &=& \kv (\ku+\kl)-\ku \kl,\\
P_{ud} &=& \kv (\ku+\kd)-\ku \kd,\\
P_{dl} &=& \kv (\kd+\kl)-\kd \kl.
\eea
If additional singlets intermix with the Higgs doublet responsible for electroweak symmetry breaking, $P_{ij} = \xi < 1$, where $\xi$ parameterizes the level of singlet mixing~\cite{Barger:2009me}.  However, in the case of additional SU(2) doublets that do not partake in fermion mass generation, the value of $\xi$ can exceed unity.  In Fig.~\ref{fig:pattern}, we show the pattern relations $P_{ul}$ for the general model. Near the SM limit, the pattern relations are highly correlated.  Therefore, we only show $P_{ul}$. 


\begin{figure}[!t]
\begin{center}
     \includegraphics[angle=0,width=0.47\textwidth]{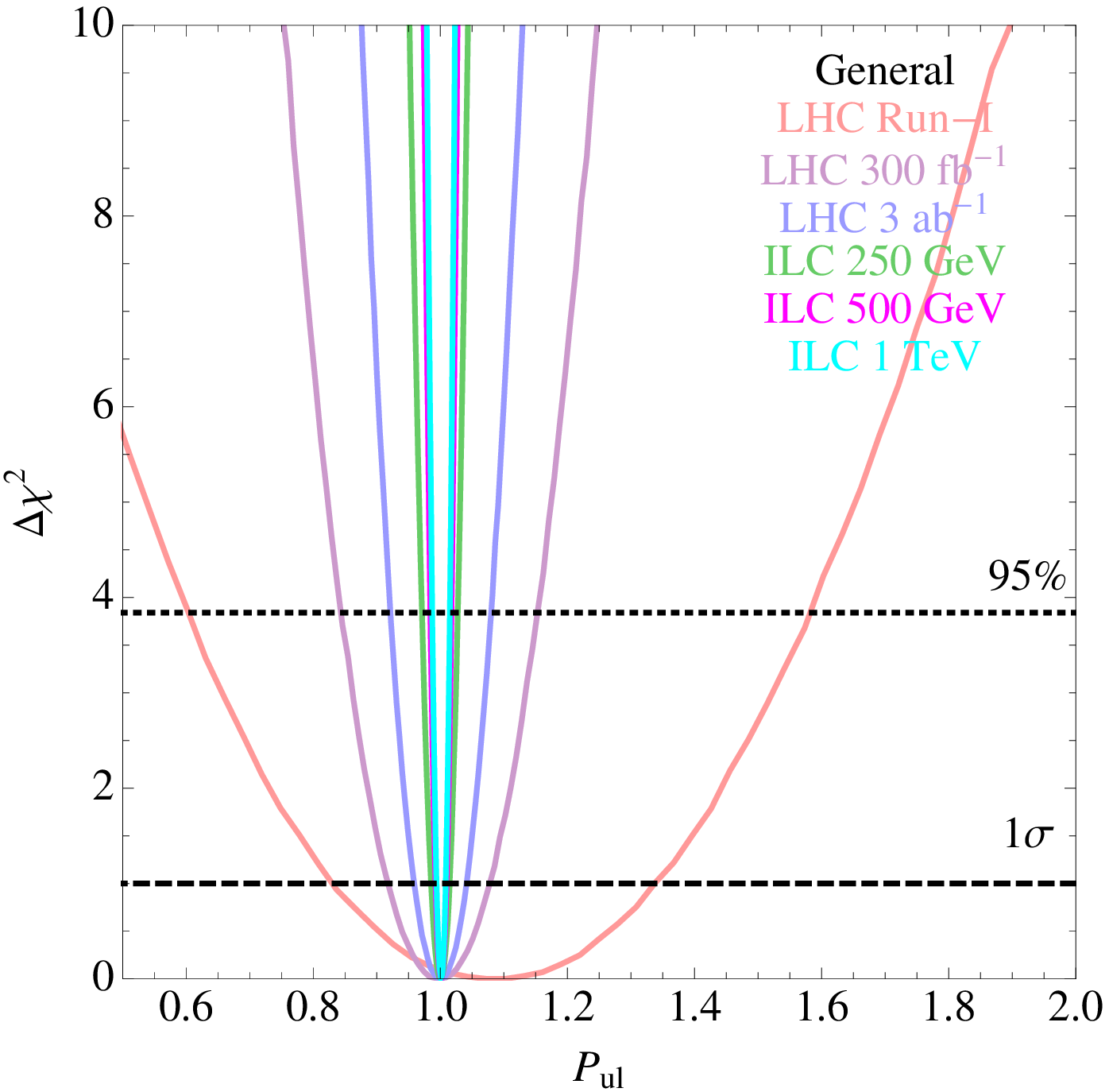}
     \includegraphics[angle=0,width=0.47\textwidth]{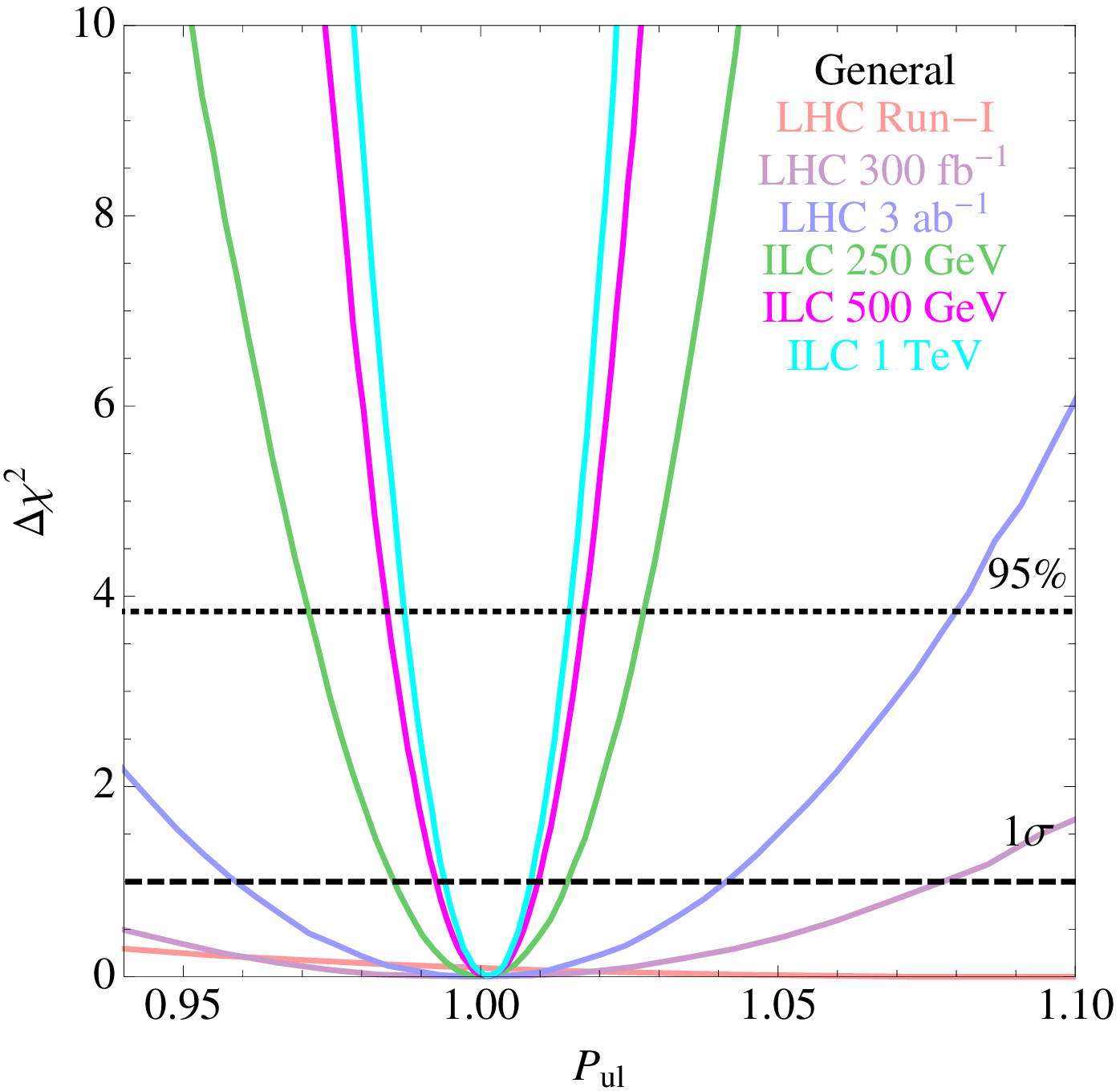}
\caption{Pattern relations in the general model. While the LHC can distinguish at the 95\% C.L. various levels of singlet and doublet mixture to 20\% with 300 fb$^{-1}$, the ILC can probe the pattern relations to the percent level. The right panel is a zoom to clarify the ILC sensitivity.}
\label{fig:pattern}
\end{center}
\end{figure}

\subsection{Singlets}
Assume a singlet scalar, $S$, mixes with Higgs doublets such that
\be
h=\cos\theta~ h^\prime + \sin \theta~ S,
\ee
where $h^\prime = \cos\alpha\, \phi_u - \sin\alpha\, \phi_d$ corresponds to the lightest Higgs boson in a 2HDM.  As the $SU(2)$ content of the Higgs boson is suppressed, the couplings to SM fields, and therefore the pattern relations, are reduced by a common factor, $\xi=\cos^2\theta$.  This reduction is in addition to the usual $\phi_u-\phi_d$ mixing (and therefore the reduction for gauge bosons $\kv=\cos(\beta-\alpha)$).  The reaches for various collider scenarios are given in Fig.~\ref{fig:singlet}.  The sensitivity to $\sin\theta$ is slightly worse than $\cos(\beta-\alpha)$ for the ILC.  This is due to the measurement of $\cos(\beta-\alpha)$ coming from the measurement of relative rates, while the $\sin\theta$ parameter controls the absolute event rate.  Therefore, only the ILC $Zh$ cross section measurement can limit the value of $\sin\theta$, whereas all other measurement help limit the value of $\cos(\beta-\alpha)$.  For example, the ILC operating at 250 GeV would, restrict $\sin\theta \lesssim 0.14$ or $\cos(\beta-\alpha)\lesssim 0.1$ at the 95\% C.L..
\begin{figure}[h!]
\begin{center}
     \includegraphics[angle=0,width=0.6\textwidth]{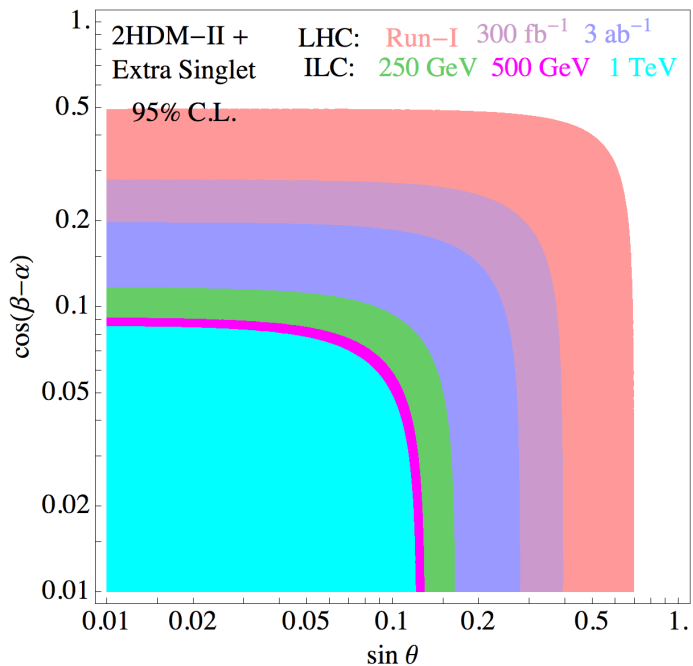}
\caption{(a) Regions in $\cos(\beta-\alpha)=0$ and $\sin\theta$ that are consistent with SM expectations with the listed collider benchmarks.   Outside the respective colored regions are areas that may be probed for the existence of additional singlets.}
\label{fig:singlet}
\end{center}
\end{figure}

\subsection{Doublets}
A 2HDM mixing with a third scalar doublet that does not couple to fermions poses an interesting scenario.  Generally, the pattern relation $P_{ij}$ can be greater or less than 1.  The observed Higgs boson can be written as
\be
h=\cos\theta~ h^\prime + \sin \theta~ \phi_0,
\ee
where $\phi_0$ is the $SU(2)$ doublet that does not couple to fermions, but may participate in EWSB.  As a consequence, the new doublet may contain an extra VEV, $\langle \phi_0 \rangle = v_0 = \sin \Omega ~ v_{\rm SM}$, that contributes to $W$ and $Z$ boson masses.  The couplings then take the form
\bea
\kv&=&\cos \theta \cos \Omega \sin(\beta-\alpha) + \sin \theta\sin\Omega,\\
\ku&=&\zeta \cos \alpha /\sin\beta,\\
\kd&=&-\zeta\sin(\alpha-\gamma_d)/\cos(\beta-\gamma_d),\\
\kl&=&-\zeta\sin(\alpha-\gamma_\ell)/\cos(\beta-\gamma_\ell),
\eea
where
\be
\zeta = \cos\theta/\cos\Omega
\ee
is  a common scale factor for all Yukawa couplings.
\begin{figure}[t!]
\begin{center}
     \includegraphics[angle=0,width=0.47\textwidth]{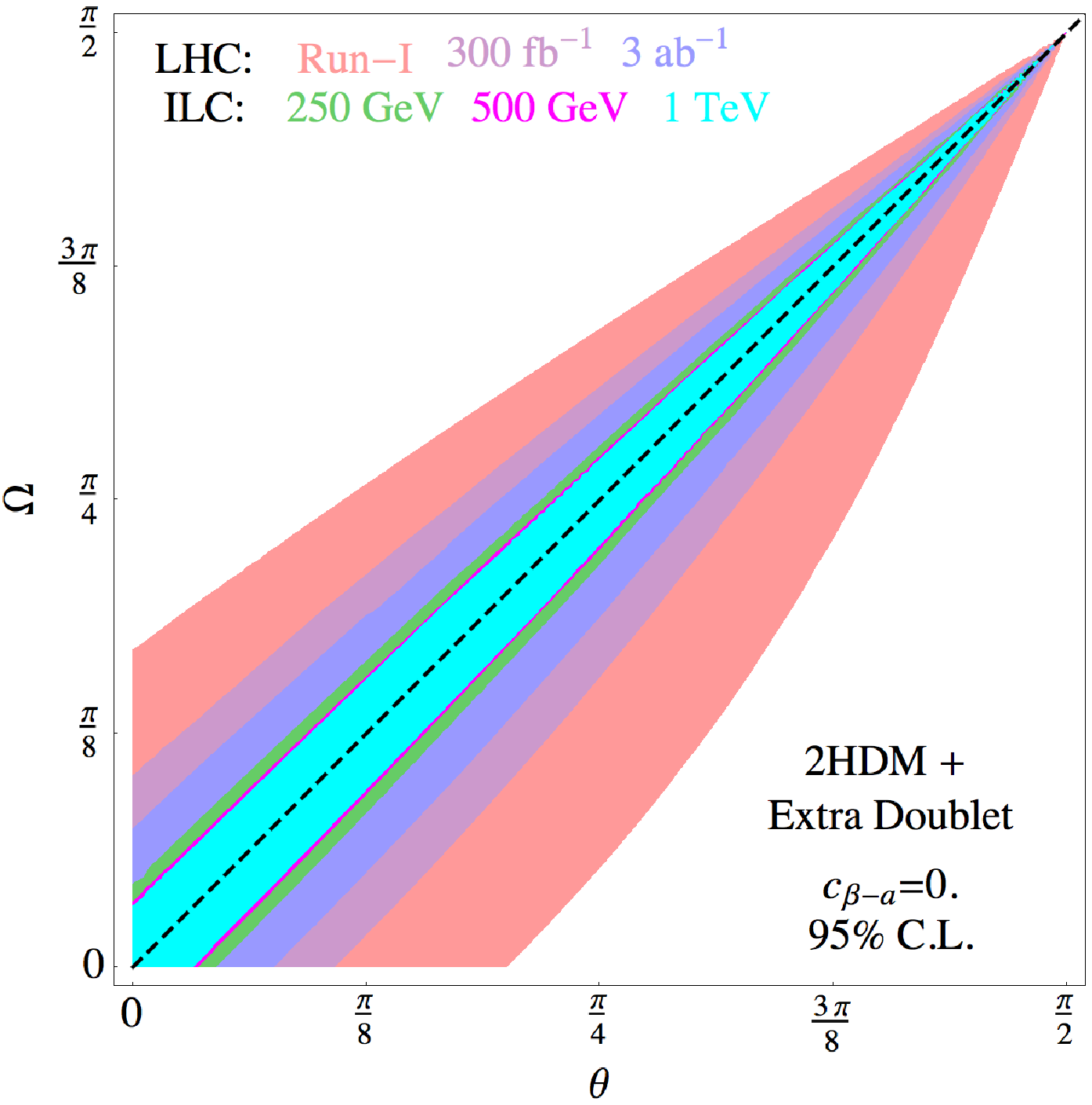}
     \includegraphics[angle=0,width=0.47\textwidth]{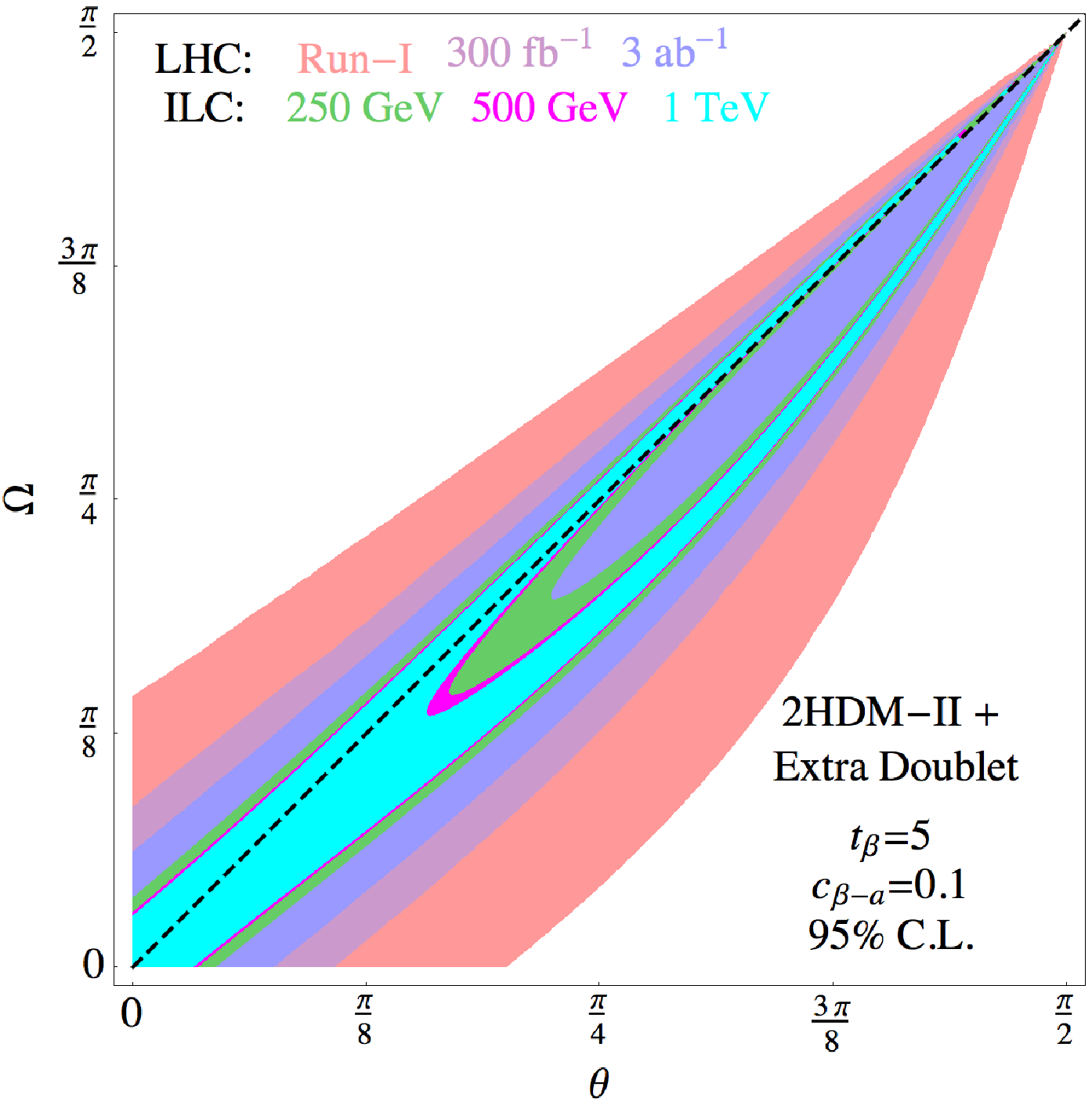}\\
     \includegraphics[angle=0,width=0.47\textwidth]{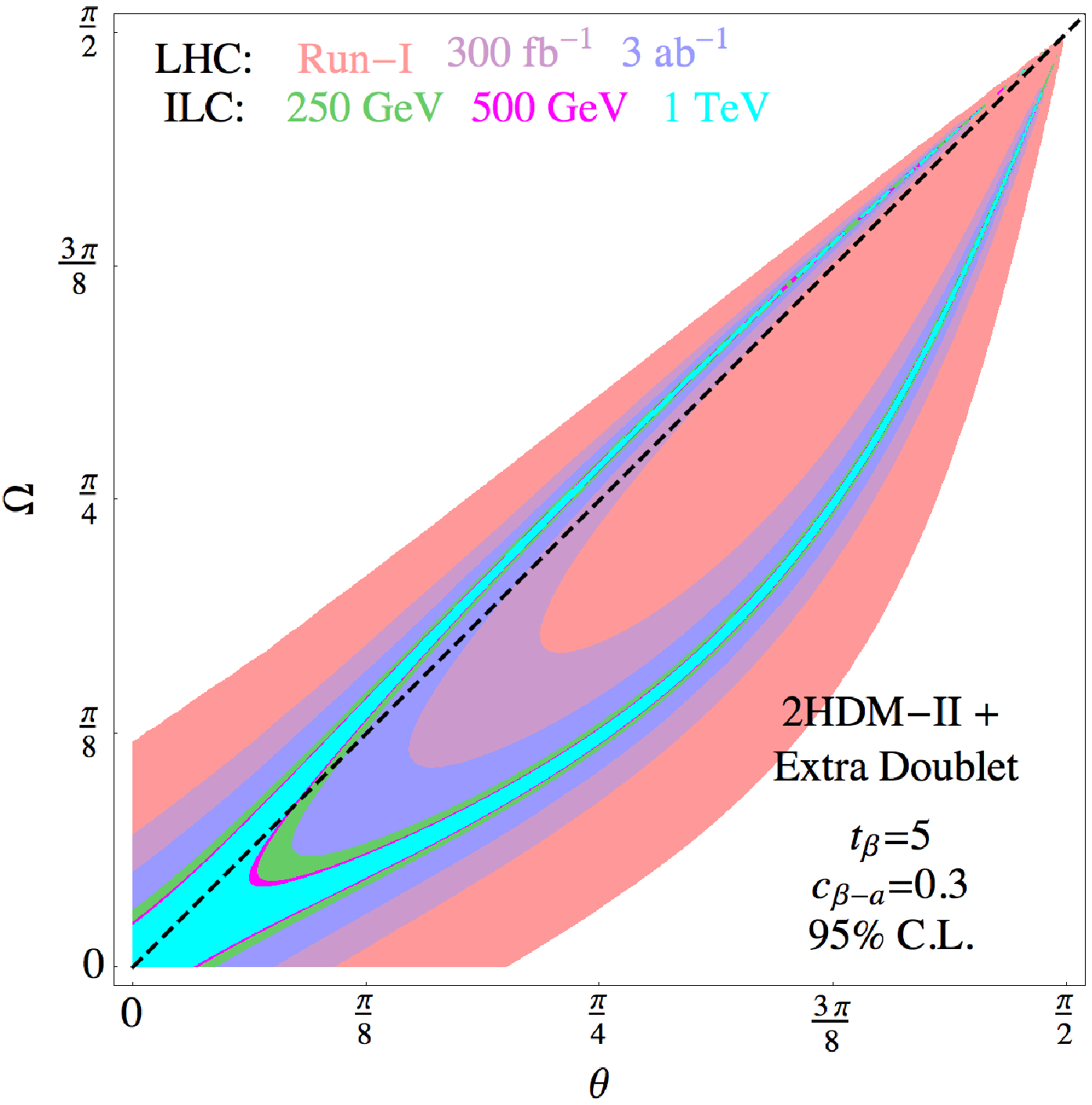}
     \includegraphics[angle=0,width=0.47\textwidth]{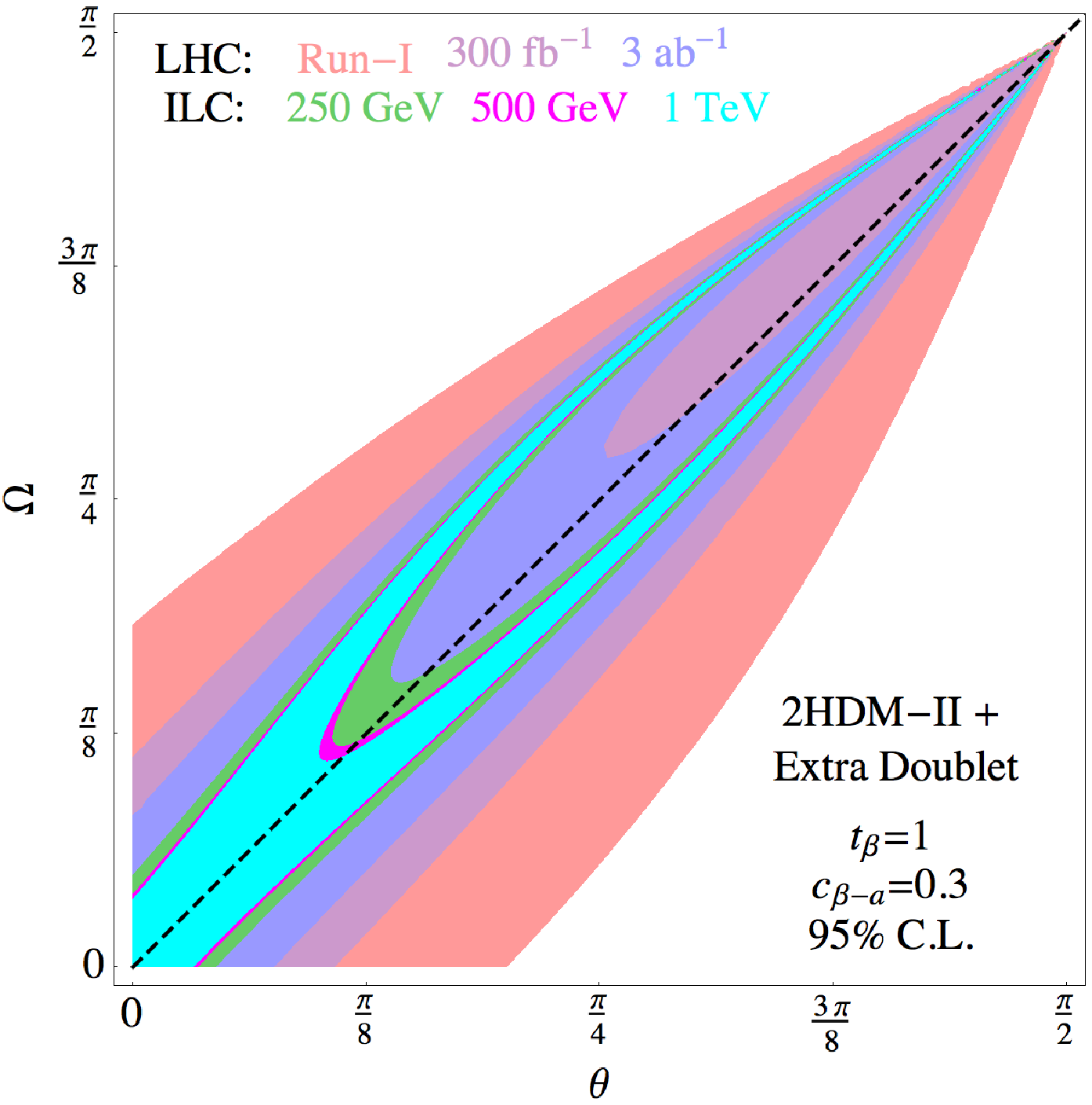}
\caption{(a) Regions in $\Omega$ and $\theta$ that are consistent with the SM central values at the decoupling limit, $\cos(\beta-\alpha)=0$.  (b-d) Examples in the 2HDM-II model away from the decoupling limit for selected values of $\cos(\beta-\alpha)$ and $\tan \beta$.  Outside the respective colored regions are areas that may be probed for the existence of additional doublets.  The diagonal dashed line marks $\Omega = \theta$.}
\label{fig:OmegaTheta}
\end{center}
\end{figure}

The 2HDM limit is retained when $\theta,\Omega\to 0$.  Note that the SM limit also occurs when $\theta \to \Omega$ and $\delta\to 0$.  Near the 2HDM and decoupling limits, one can expand the pattern relations about $\delta$ and $\Delta_{\Omega\theta}=\Omega-\theta$ to second order,
\bea
P_{ul} &\sim& 1 - \delta^2\left({t_{\gamma_\ell-\beta}\over t_\beta}+c^2_\theta\right) - \delta \Delta_{\Omega\theta}{ t_\theta c_{\gamma_\ell-2\beta}\over s_\beta c_{\gamma_\ell-\beta}} -{\Delta_{\Omega\theta}^2\over c_\theta^2} 
\label{eq:patternexpstart}
\\
P_{ud} &\sim&1 - \delta^2\left({t_{\gamma_d-\beta}\over t_\beta}+c_\theta^2\right) - \delta \Delta_{\Omega\theta}{ t_ \theta c_{\gamma_d-2\beta}\over s_\beta c_{\gamma_d-\beta}} -{\Delta_{\Omega\theta}^2\over c_\theta^2} \\
P_{dl} &\sim&1 - \delta^2\left(t_{\gamma_\ell-\beta} t_{\gamma_d-\beta}+c_\theta^2\right) - \delta \Delta_{\Omega\theta}{ t_\theta s_{\gamma_d+\gamma_\ell-2\beta}\over c_{\gamma_d-\beta} c_{\gamma_\ell-\beta}} -{\Delta_{\Omega\theta}^2\over c_\theta^2},
\label{eq:patternexpend}
\eea
where the trigonometric functions are denoted by $s, c, t$ with argument specified by the subscript.  The decoupling line can be seen in Fig.~\ref{fig:OmegaTheta}a for $\cos(\beta-\alpha)\equiv \delta = 0$, the shaded regions correspond to machine benchmark sensitivities with data assumed to have SM central values.  In this instance, the choice of $\gamma_d$ and $\gamma_\ell$ do not impact the reach in the $\theta, \Omega$ plane. However, once $\cos(\beta-\alpha)\ne 0$, the choice of $\gamma_{d,\ell}$ and $\tan\beta$ alter the contours.  In Fig.~\ref{fig:OmegaTheta}(b-d), we show these cases for select parameters within the 2HDM-II with an extra doublet.  Similar features are present for the other models.  The bifurcation of the contours in these cases is driven by the $\delta \Delta_{\Omega\theta}$ term in Eqs.~(\ref{eq:patternexpstart}-\ref{eq:patternexpend}).  For sufficiently large values of $\tan\beta$, the ${\cal O}(\delta \Delta_{\Omega\theta})$ term dominates and provides a shift 
\be
 \delta \Delta_{\Omega\theta}{ t_ \theta c_{\gamma_d-2\beta}\over s_\beta c_{\gamma_d-\beta}} \to \delta \Delta_{\Omega\theta}\tan\theta\tan\beta,
 \ee
which shifts the pattern relation values and leads to the excluded regions along the $\Omega = \theta$ line.  Note in the $\tan\beta=1$ case (Fig.~\ref{fig:OmegaTheta}d), higher order terms dominate.  It is also worth noting that an excluded region of moderate $\theta=\Omega$ is possible at the ILC and, depending on the parameter choices of $\tan\beta$ and $\cos(\beta-\alpha)$, the LHC.  Therefore, the existence of additional doublets may be probed at LHC14 and ILC.

\section{Discussion and Summary}
\label{sect:concl}
We make the following projections for distinguishing 2HDMs from the SM with LHC, ILC or Muon Collider data.  The numbers quoted below are based on a 1d parameter $\Delta\chi^2$ fit, while the figures referenced below correspond to a 2d parameter fit unless otherwise specified.\footnote{Recall that the $\Delta \chi^2$ fits for a 1d parameter fit require $\Delta \chi^2=1$, while for a 2d fit $\Delta \chi^2=2.30$. }  We summarize the capabilities of each machine benchmark in Fig.~\ref{fig:barfigs} for the general model, the decoupling value $\cos (\beta-\alpha)$ in 2HDMs, and a metric we call the ``Distinguishing Power''.  We define the Distinguishing Power as 
\be
DP = \sqrt{A_{1\sigma}\over \pi},
\label{eq:distpwr}
\ee  
where $A_{1 \sigma}$ is the area of the $1\sigma$ uncertainty region in a given parameter plane.  The distinguishing power encapsulates the total uncertainty in the parameter plane and is related to the geometric mean of the principal axes of an error ellipse, but is more generally applicable.  Note that while this metric does not retain correlation information, it serves as a figure of merit when comparing the capabilities of various machine benchmarks.  

\begin{figure}[]
\begin{center}
     \includegraphics[angle=0,width=0.80\textwidth]{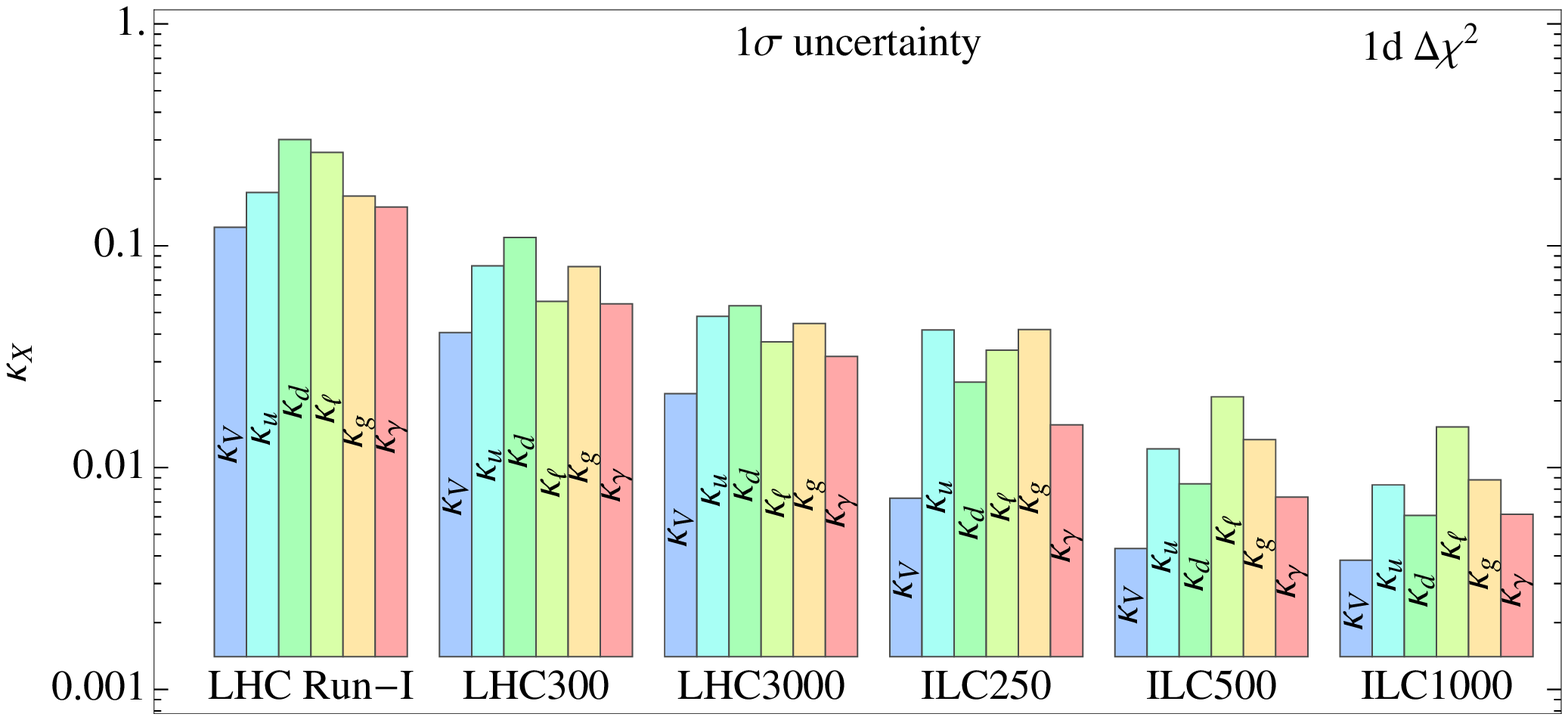}
     \includegraphics[angle=0,width=0.80\textwidth]{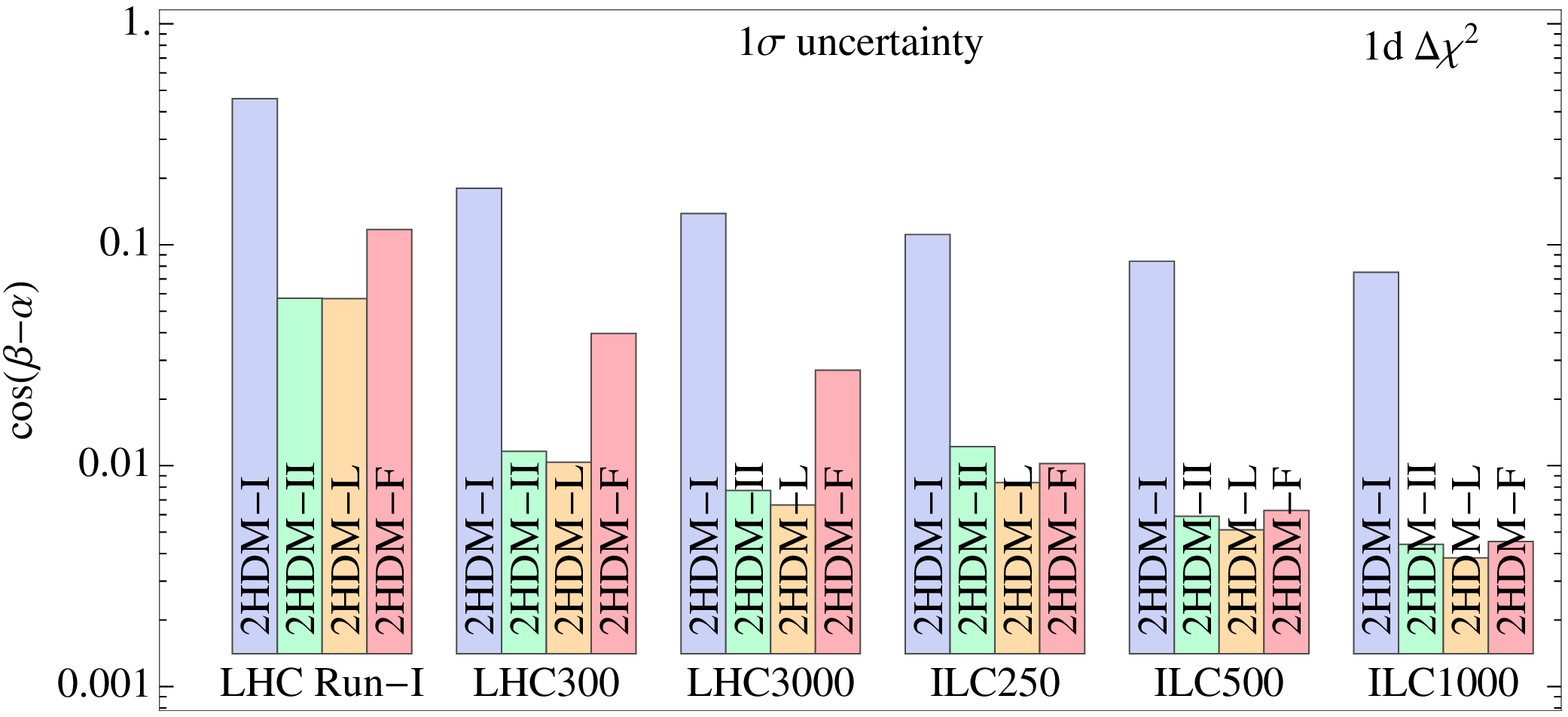}
     \includegraphics[angle=0,width=0.80\textwidth]{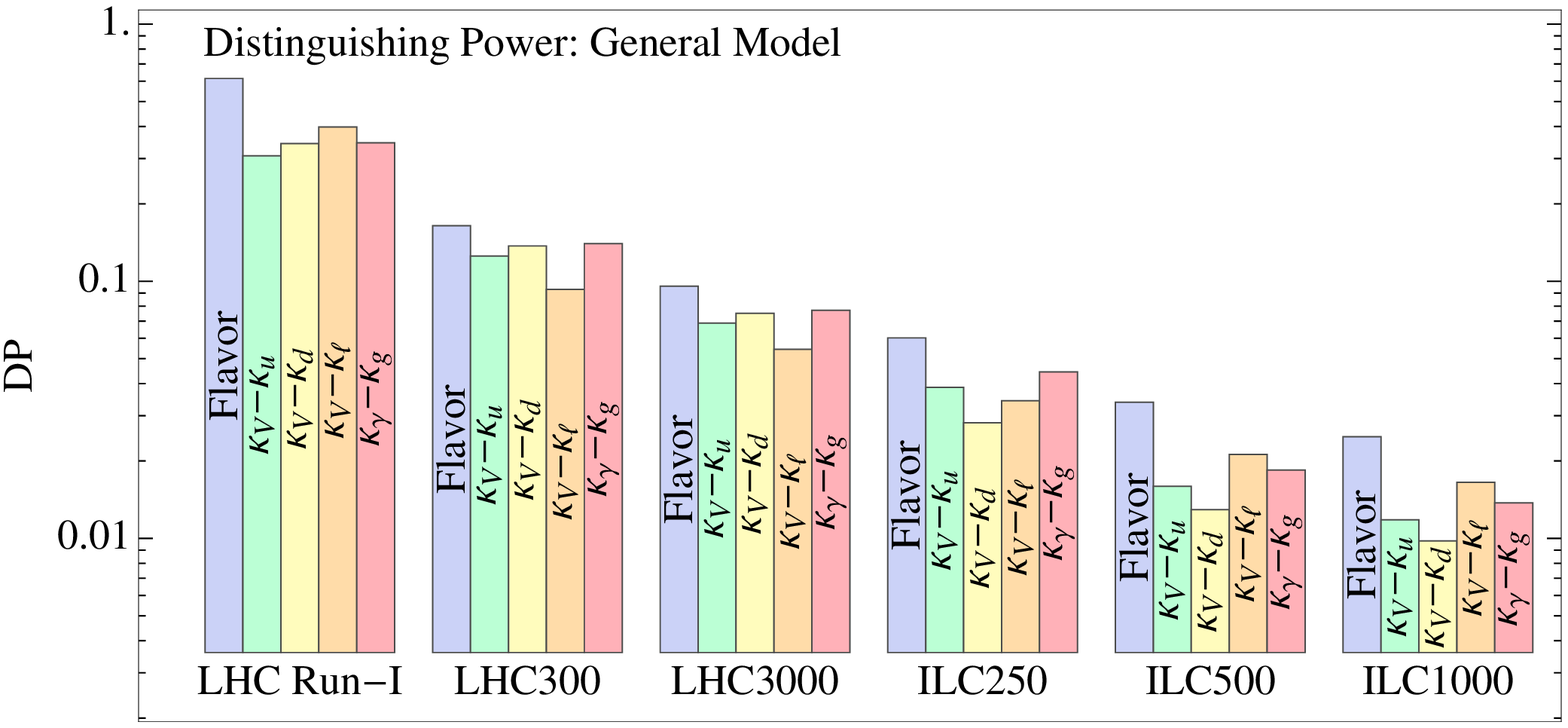}
\caption{Comparison at the $1\sigma$ level of machine benchmarks for coupling measurements in the general model assuming no new physics decays (top panel) and the largest deviation of $\cos(\beta-\alpha)$ from zero in 2HDMs (middle panel).  The distinguishing power, $DP$, for selected parameter planes in the general model (bottom panel).  The flavor $DP$ within the $\kd/\ku$ vs. $\kl/\ku$ plane is given in Fig.~\ref{fig:genfitdiff}.}
\label{fig:barfigs}
\end{center}
\end{figure}

\bi
\item The measurement of Yukawa coupling ratios tests the Higgs flavor structure.  Specifically, the LHC at 300 \ifb (3 \iab) can measure the coupling ratios of $\kd/\ku$ and $\kl/\ku$ to  6\% and 9\% (3\% and 7\%).  The ILC  250, 500, 1000 can measure these ratios to 4\%, 2\% and 1.5\%, respectively.  Fig.~\ref{fig:genfitdiff} gives the regions consistent with SM values in the planes of $\kd/\ku$ and $\kl/\ku$.  The distinguishing power, $DP$, between two machine benchmarks is related to the geometric mean of the principal axes length in the $1\sigma$ error region for a specific parameter plane (c.f. Eq.~\ref{eq:distpwr}). For flavor non-universal couplings (i.e. in the $\kd/\ku$ vs. $\kl/\ku$ plane in Fig.~\ref{fig:genfitdiff}a) with the machine benchmarks we consider, it proceeds in the successive order: $DP= $ 0.35, 0.10, 0.059, 0.037, 0.021, and 0.015 for the LHC Run-I, LHC300, LHC3000, ILC250, ILC500, and ILC1000, respectively.  The values of $DP$ for various parameter planes are shown in Fig.~\ref{fig:barfigs}c.

\item The ``wrong Higgs'' couplings, those which exist only at the loop level and are induced via loops that may include new physics, can occur in 2HDMs. These are manifest as shifts in the $b$ and $\tau$ Yukawa couplings and can be easily probed at a future ILC.  Fig.~\ref{fig:delb} shows the effects of the shift in the $b$ Yukawa in the $\kd/\ku$ and $\kl/\ku$ plane for the machine benchmarks.

\item The total Higgs width can be inferred at the LHC if we restrict $\kv < 1$ to be consistent with unitarity in models containing doublets and/or singlets as shown in Fig.~\ref{fig:width}b.  The ILC250 can measure the total $Zh$ production cross section, thereby fixing the normalization of $\kv$ and thus constraining the total Higgs width: see Fig.~\ref{fig:width}c.  This amounts to being sensitive to new physics contributions to the Higgs decay of  $\Gamma_X \gtrsim 0.25 $ MeV at the 95\% C.L..  

\item A Muon Collider scan over the Higgs resonance profile is the only way to directly measure the total Higgs width.  This is anticipated to yield  a $< 4\%$  uncertainty in the width~\cite{Han:2012rb}.    In turn, this provides sensitivity to new physics decays contributions of $\Gamma_X \gtrsim 0.11 $ MeV at the 95\% C.L. (c.f. Fig.~\ref{fig:width}c).

\item Since the $h\to b\bar b$ rate is a substantial portion of the total Higgs decay width, an improvement in the $\kd$ measurement can be made at a Muon Collider operating on the Higgs resonance.  To a lesser extent this is true of the $\kl$ extraction as well.

\item Consistency of the Higgs boson couplings with the measured $W$ and $Z$ boson masses can be probed to remarkable precision.  In the Type-II, L and F models, the vector boson coupling can be measured down to $\kv = \cos(\beta-\alpha) \sim 0.4\%$ at ILC1000, via the Vector Boson coupling measurements.  For the Type-I model, it can be measured down to 8\%.

\item Additional singlets reduce the gauge and Yukawa couplings, while extra doublets can reduce or enhance Yukawa couplings.  Pattern relations among the gauge and Yukawa couplings can help establish the underlying model~\cite{Barger:2009me}.  Therefore, the future LHC and ILC data are well positioned to further probe and distinguish these mixing scenarios. The singlet mixing can be constrained to the $\sin\theta \sim {\cal O}(0.1)$ level with the ILC1000 benchmark, (c.f. Fig.~\ref{fig:singlet}).

\ei


\section{Acknowledgements}
V.B. and L.L.E. thank KITP for hospitality during a portion of this work.  V.B. thanks the Physics Department of the University of Hawaii at Manoa for hospitality.  V.B., L.L.E. and G.S. are supported by the U. S. Department of Energy under the contract DE-FG-02-95ER40896.  H.E.L. is supported by the Natural Sciences and Engineering Research Council of Canada.

\appendix
\section{ Higgs Measurements}
\label{apx:CurrentHdata}
The combined CDF and D0 data on inclusive Higgs production, based on 10 fb$^{-1}$ integrated luminosity, give the best fit values in Table~\ref{tab:FNAL} at a Higgs mass of 125 GeV. The CMS and ATLAS data on Higgs signals with integrated luminosity of $17.0-25.5$ fb$^{-1}$ lead to the best fit values summarized in Tables~\ref{tab:CMS} and \ref{tab:ATLAS}.

\begin{table}[!h]
\caption{Best fit values of the Higgs cross sections relative to the SM prediction at CDF and D0 for $M_h=125$ GeV~\cite{Tevatron:2012cn}.  Uncertainties quoted are $1\sigma$.}
\begin{center}
\begin{tabular}{|c|c|}
\hline
Channel & $\mu$\\
\hline
$p\bar p\to H\to WW$ & $0.32^{+1.13}_{-0.32}$\\
$VH$, $H\to b\bar b$ & $1.97^{+0.74}_{-0.68}$\\
$p\bar p\to H\to \gamma\gamma$ & $3.62^{+2.96}_{-2.54}$\\
\hline
\end{tabular}
\end{center}
\label{tab:FNAL}
\end{table}%
\begin{table}[!h]
\caption{Best fit values of the production and decay Higgs cross sections relative to the SM prediction at CMS for $M_h$ near 125 GeV.  Mass assumptions vary by channel, but the large uncertainties currently associated with the cross sections are not expected to have an appreciable impact on the fit.  Uncertainties quoted are $1\sigma$.}
\begin{center}
\begin{tabular}{|c|c|c|c|}
\hline
Channel & $\mu$ & ${\cal L} dt$ (fb$^{-1}$) & Reference\\
\hline
$\gamma\gamma$ 0,1-jet & $0.70^{+0.33}_{-0.29}$& 24.7 &  \cite{CMS:yva,CMS:ril}\\
$\gamma\gamma$ VBF-tag & $1.01^{+0.63}_{-0.54}$& 24.7 & \cite{CMS:yva,CMS:ril}\\
$\gamma\gamma$ VH-tag & $0.57^{+1.34}_{-1.14}$& 19.6 & \cite{CMS:yva,CMS:ril}\\
$\gamma\gamma+t\bar t$ & $-0.2^{+2.4}_{-1.9}$& 19.6 & \cite{CMS:13015}\\
\hline
$WW$ 0,1-jet & $0.73^{+0.22}_{-0.20}$& 24.4 & \cite{CMS:yva,CMS:bxa}\\
$WW$ VBF-tag & $-0.05^{+0.75}_{-0.56}$& 17.0 & \cite{CMS:yva,CMS:eya}\\
$WW$ VH-tag & $0.51^{+1.26}_{-0.94}$& 24.4 & \cite{CMS:yva,CMS:zwa}\\
\hline
$ZZ$ 0,1-jet & $0.86^{+0.32}_{-0.26}$& 24.7 & \cite{CMS:yva,CMS:xwa}\\
$ZZ$ VBF-tag & $1.24^{+0.85}_{-0.58}$& 24.7 & \cite{CMS:yva,CMS:xwa}\\
\hline
$\tau\tau$ 0,1-jet & $0.77^{+0.58}_{-0.55}$& 24.5 & \cite{CMS:yva,CMS:utj}\\
$\tau\tau$ VBF-tag & $1.42^{+0.70}_{-0.64}$& 24.5 & \cite{CMS:yva,CMS:utj}\\
$\tau\tau$ VH-tag & $0.98^{+1.68}_{-1.50}$& 24.5 & \cite{CMS:yva,CMS:ckv}\\
\hline
$b\bar b$ VBF-tag & $0.7^{+1.4}_{-1.4}$& 19.0 & \cite{CMS:13011}\\
$b\bar b$ VH-tag & $1.0^{+0.5}_{-0.5}$& 24.0 & \cite{CMS:13012}\\
\hline
\end{tabular}
\end{center}
\label{tab:CMS}
\end{table}%
\begin{table}[!h]
\caption{Best fit values of the production and decay Higgs cross sections relative to the SM prediction at ATLAS for $M_h$ near 125 GeV.  Mass assumptions vary by channel, but the large uncertainties currently associated with the cross sections are not expected to have an appreciable impact on the fit.  Uncertainties quoted are $1\sigma$.}
\begin{center}
\begin{tabular}{|c|c|c|c|}
\hline
Channel & $\mu$& ${\cal L} dt$ (fb$^{-1}$) & Reference\\
\hline
$\gamma\gamma$ $ggF + t\bar t H$ & $1.6^{+0.42}_{-0.36}$& 25.5 & \cite{ATLAS:2013oma}\\
$\gamma\gamma$ VBF & $1.7^{+0.94}_{-0.89}$& 25.5 & \cite{ATLAS:2013oma}\\
$\gamma\gamma$ VH & $1.8^{+1.5}_{-1.3}$& 25.5 & \cite{ATLAS:2013oma}\\
\hline
$WW$ $ggF$ & $0.82^{+0.36}_{-0.36}$& 25.3 & \cite{ATLAS:2013wla}\\
$WW$ VBF & $1.66^{+0.79}_{-0.79}$& 25.3 & \cite{ATLAS:2013wla}\\
\hline
$ZZ$ $ggF$ & $1.8^{+0.8}_{-0.5}$& 25.3 & \cite{ATLAS:2013nma}\\
$ZZ$  VBF & $1.2^{+3.8}_{-1.4}$& 25.3 & \cite{ATLAS:2013nma}\\
\hline
$\tau\tau$ $ggF$ & $2.4^{+2.4}_{-2.3}$& 17.6 & \cite{ATLAS:2012dsy}\\
$\tau\tau$ VBF & $-0.4^{+2.3}_{-1.4}$& 17.6 & \cite{ATLAS:2012dsy}\\
\hline
$b\bar b$ VH & $-0.4^{+1.1}_{-1.1}$& 17.7 & \cite{ATLAS:2012wma}\\
\hline
\end{tabular}
\end{center}
\label{tab:ATLAS}
\end{table}%

\end{document}